%% file: main.tex
\PassOptionsToPackage{table}{xcolor}
\documentclass[10pt,conference]{IEEEtran}

\input{header}

\begin{document}
\pagestyle{plain}

\title{Adversarial Trust Poisoning in Vehicular Collaborative Perception}

\author{
\IEEEauthorblockN{Yutong Liu \quad Chenyi Wang \quad Ming F. Li \quad Qingzhao Zhang}
\IEEEauthorblockA{
School of ECSE, The University of Arizona, Tucson, AZ\\
\{yutongl, chenyw, lim, qzzhang\}@arizona.edu
}
}

\maketitle
\thispagestyle{plain}
\input{abstract}

\input{introduction}
\input{related_work}
\input{threat_model}
\input{design}
\input{evaluation}
\input{discussion}
\input{conclusion}

\section*{Ethical Considerations}

This work studies security risks in collaborative perception and is inherently dual-use, since physical adversarial objects that induce trust-poisoning failures could be misused if deployed in real traffic. We therefore designed the study to minimize risk and focus exclusively on defensive security analysis.

Our evaluation is conducted primarily in digital simulation using the OPV2V and V2X-Real research platform and open-source collaborative-perception models and defenses. The work does not interact with deployed V2X infrastructure, production vehicles, or public transportation systems, and it does not intercept, modify, or transmit real V2X communication. No human-subject data or personally identifiable information is collected or analyzed.

The physical experiments are limited to sensor-level validation of LiDAR behavior using physically constructed prototypes. These experiments were performed under controlled conditions on a closed and unoccupied road segment. All vehicles remained stationary during data collection, no human participant operated a vehicle during the experiments, and no attack was evaluated in live traffic or against public infrastructure.

To reduce misuse risk, we evaluate only research collaborative-perception pipelines and published defenses rather than targeting vendor-specific deployed systems. We additionally present \defensename as a mitigation and release artifacts to support reproducible evaluation and future defensive research. We believe that responsible publication is justified because the work identifies a previously underexplored blind spot in consistency-based defenses and highlights the need for more robust trust mechanisms in collaborative perception systems.

\bibliographystyle{IEEEtran}
\bibliography{references}

\appendix
\input{appendix}

\end{document}

%% file: header.tex
\usepackage{algorithm}
\usepackage{algpseudocode}
\usepackage{soul}
\usepackage{svg}
\usepackage{graphicx}
\usepackage{mwe}  
\usepackage{filecontents}
\usepackage{diagbox}

\usepackage{tikz}
\usepackage{amsmath}

\usepackage{filecontents}

\usepackage{booktabs}    
\usepackage{multirow}    
\usepackage{caption}
\usepackage{subcaption}
\usepackage{tabularx}

\usepackage{xurl} 
\usepackage{amssymb}

\usepackage{colortbl}
\usepackage{array}
\usepackage{threeparttable}
\usepackage{gensymb}
\usepackage{xcolor}
\usepackage[table]{xcolor}
\definecolor{Gray}{gray}{0.9}
\definecolor{LightGreen}{RGB}{230,255,230}
\usepackage{xspace}
\usepackage{siunitx}
\usepackage{placeins}

\usepackage{tcolorbox}
\tcbuselibrary{breakable,skins}  

\usepackage{hyperref}
\usepackage{makecell}

\newcommand{\attackname}{\texttt{TrustFlip}\xspace}
\newcommand{\defensename}{\texttt{TrustReflect}\xspace}
\newcommand{\fromreal}{\textsc{FromReal}\xspace}

\usepackage{enumitem}
\setlist[itemize]{nosep, itemsep=1pt, topsep=1pt, leftmargin=1em, labelwidth=*}
\setlist[enumerate]{nosep, itemsep=1pt, topsep=1pt, leftmargin=1em, labelwidth=*}

\newcommand{\myparagraph}[1]{\smallskip\noindent\textbf{#1}:\xspace}

\newcounter{challengecnt}
\newcounter{insightcnt}[challengecnt]
\renewcommand{\thechallengecnt}{\arabic{challengecnt}}
\renewcommand{\theinsightcnt}{\thechallengecnt.\arabic{insightcnt}}
\newcommand{\challenge}[1]{%
  \refstepcounter{challengecnt}%
  \smallskip\noindent\textbf{Challenge \thechallengecnt: #1}.\xspace}
\newcommand{\insight}[1]{%
  \refstepcounter{insightcnt}%
  \par\smallskip\noindent\textbf{Insight \theinsightcnt.} \emph{#1}\xspace}

\usepackage[textsize=tiny,colorinlistoftodos,textwidth=15mm]{todonotes} 

\newtcolorbox[auto counter]{finding}[1][]{
  colback=black!5!white, colframe=black!75!black, boxrule=0.8pt, left=2pt, right=2pt, top=2pt, bottom=2pt,
  enhanced, drop shadow,
  breakable,
  fonttitle=\scshape,
  before upper=\textbf{Finding \thetcbcounter:\ },
  #1
}

%% file: abstract.tex
\begin{abstract}
Collaborative perception (CP) enables connected and autonomous vehicles to share sensor data and jointly reason about their environment. To defend against adversaries that fabricate or manipulate shared data, existing systems employ cross-vehicle inconsistency detection and trust estimation, penalizing vehicles whose observations conflict with the majority. In this work, we show that these defenses themselves introduce a new attack surface. We present \attackname, a novel attack that weaponizes consistency-based defenses to poison the trust assigned to benign vehicles. Instead of injecting false data into the collaboration pipeline, it deploys physical adversarial objects that induce inconsistent observations among benign vehicles. The resulting inconsistencies are misattributed by the defense to the targeted victim, causing its trust score to degrade and eventually leading to its downweighting or exclusion from collaboration. Consequently, the system loses reliable sensing contributors, degrading perception capability and potentially inducing safety-critical failures. On the OPV2V and V2X-Real benchmark, \attackname causes CP defenses to flag or suppress the targeted benign vehicle in up to 87.76\% of scenarios and reduces average precision (AP) by up to 6.0 percentage points. Real-world prototypes of such adversarial objects further validate the attack practicality. As an initial mitigation to this new type of attack, we further propose \defensename, a lightweight self-reflection mechanism that excludes locally disputed regions from trust evaluation and reduces attack success rate by 35--100\% on the same benchmark.
\end{abstract}

%% file: introduction.tex
\section{Introduction}
\label{sec:introduction}

Connected and Autonomous Vehicles (CAVs) are transforming transportation through enhanced intelligence and cooperation~\cite{CAV}. Collaborative perception (CP) is a representative application in which vehicles and roadside infrastructure share perception data (raw sensor measurements, intermediate neural features such as bird's-eye-view feature maps, or final detection outputs such as bounding boxes) to jointly perceive the environment. Prior work shows that such collaboration substantially improves sensing range and reliability over single-vehicle perception, especially for blind spots and challenging objects~\cite{han2023collaborative,wang2020v2vnet,liu2020when2com,xu2022v2x,zhang2023robust}.

However, CP systems are vulnerable to data fabrication attacks, where a compromised participant shares falsified information that spoofs, removes, or perturbs objects~\cite{tu2021adversarial,zhang2024data,wang2025threat,wang2025cp,zhang2024stealthy,lin2025pretend,zhang2026stealthy}. To mitigate such threats, existing defenses employ anomaly detection, trust estimation, and recovery mechanisms built around cross-agent consistency. The intuition is that fabricated data will conflict with observations from the majority of benign vehicles~\cite{li2023among,zhang2024data,zhao2024made,wang2025threat}. ROBOSAC~\cite{li2023among} and CAD~\cite{zhang2024data} flag vehicles whose shared data (bounding boxes or occupancy maps) fall out of consensus. CP-Guard+~\cite{hu2025cp} and LUCIA~\cite{wang2025threat} analyze shared intermediate features to identify feature-level deviations from a trusted ego or the majority of vehicles. MADE~\cite{zhao2024made} compares shared features against an ego-only reconstruction and flags outliers. MATE~\cite{hallyburton2025security} observes object tracking level inconsistencies and assigns probabilistic trust scores via a Bayesian framework.

Although effective against direct data fabrication, consistency-based defenses rely on an attribution assumption: a contributor that disagrees with other vehicles is likely to be malicious. This assumption fails when benign vehicles legitimately perceive the same physical object differently from different viewpoints. We show that an attacker can deliberately create such disagreement, making genuine data from a targeted benign collaborator appear malicious without compromising any participant. We term this attack \textbf{adversarial trust poisoning}.

\begin{figure}[t]
    \centering
    \includegraphics[width=\columnwidth]{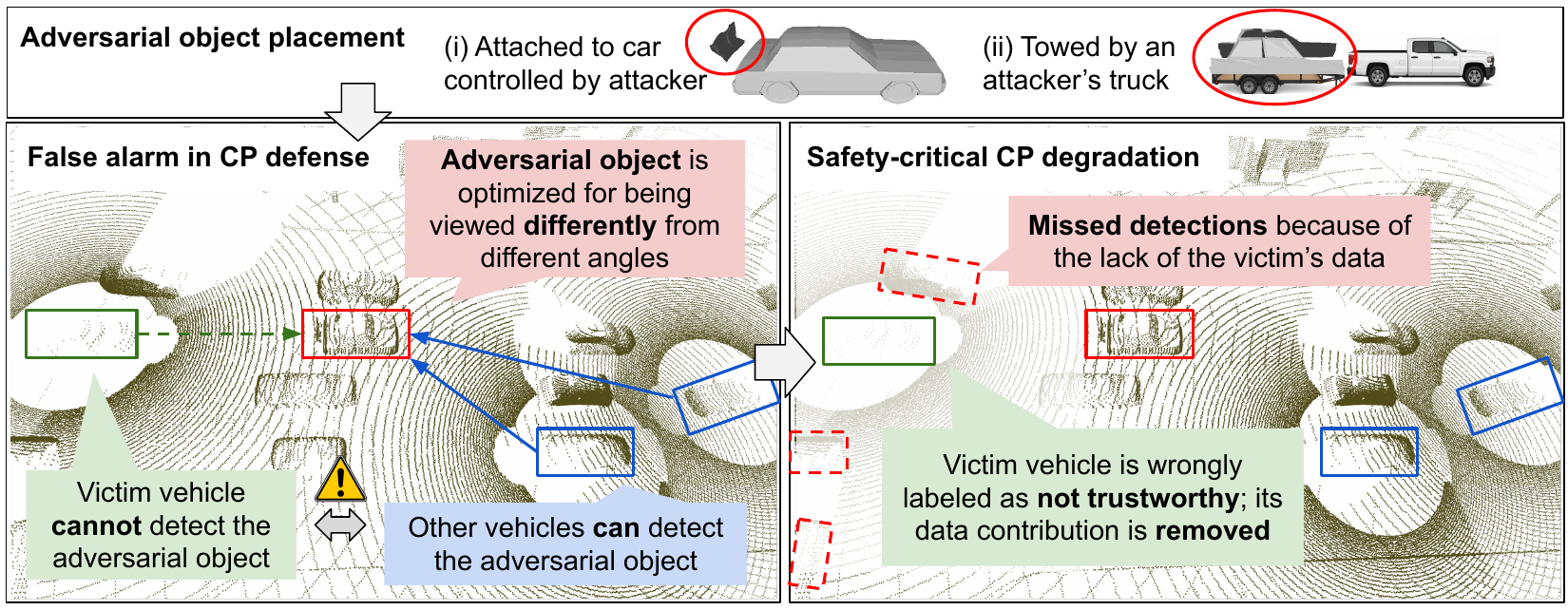}
    \caption{Demonstration of \attackname. (1) A physical adversarial object induces viewpoint-dependent perception. (2) The victim misses the object while other benign vehicles retain it. (3) The CP defense misattributes the disagreement to the victim and excludes its data, potentially causing the fused perception to miss objects that only the victim observes.}
    \label{fig:demo}
\end{figure}

We instantiate this attack as \attackname using a \textbf{physical adversarial object}. Unlike prior physical attacks whose endpoint is a detector error, it uses viewpoint-dependent perception as the mechanism for manipulating a downstream defense decision. Compared with active alternatives such as LiDAR spoofing~\cite{jin2023pla} or jamming~\cite{petit2014potential}, the object is passive and requires no signal-injection hardware, synchronization, or vehicle compromise. As shown in Figure~\ref{fig:demo}, the attacker places one object so that a designated benign CAV (the \emph{victim}) and surrounding benign CAVs obtain inconsistent observations of it. The victim's genuine perception output then conflicts with the apparent consensus. The defense identifies the victim as the source of this disagreement and downweights or excludes its contribution, degrading CP where that vehicle provides useful sensing coverage.

Realizing \attackname is challenging because the victim and non-victims often observe overlapping object surfaces: geometry changed to hide the object from one view also destroys the evidence that keeps it visible from the other, so the two view-conditioned objectives conflict on shared geometry.
\attackname addresses this challenge with three physically biased shape priors: a realistic vehicle mesh perturbed only within a constrained region, an ordinary vehicle augmented with a small rear-mounted structure, and a standalone hollow structure whose parallel panels let LiDAR rays pass from one direction while reflecting from another. Each prior physically satisfies or preserves one side of the view-conditioned objective, reducing conflict during constrained, defense-aware optimization. \attackname further incorporates runtime scenario assessment, victim selection, and object placement so that the induced disagreement reaches the defense and affects CP utility under realistic deployment conditions.

As an initial mitigation, we propose \defensename, a lightweight self-reflection layer for consistency-based defenses. Each vehicle applies the defense to its local perception and marks flagged regions as \emph{uncertain} in a per-frame \emph{trust mask}. These regions are excluded from downstream trust computation, preventing local uncertainty from being misattributed to a benign vehicle. \defensename requires no retraining and only one additional V2X message field, though ambiguity between benign and adversarial inconsistencies may remain.

Our extensive evaluation reveals a \textbf{sensitivity-attribution dilemma}: the same sensitivity that lets a consistency-based defense catch subtle data fabrication is what exposes it to trust poisoning. We establish this finding on two CP datasets, OPV2V~\cite{xu2022opv2v} and V2X-Real~\cite{xiang2024v2x}, across four state-of-the-art (SOTA) consistency-based defenses (CAD~\cite{zhang2024data}, MATE~\cite{hallyburton2025security}, LUCIA~\cite{wang2025threat}, MADE~\cite{zhao2024made}) and four CP backbones spanning late fusion (PIXOR~\cite{yang2018pixor}, PointPillars~\cite{lang2019pointpillars}) and intermediate fusion (Attentive Fusion~\cite{xu2022opv2v}, Where2Comm~\cite{hu2022where2comm}). \attackname reliably induces the intended view-conditioned asymmetry, propagates it into multiple defense mechanisms, and degrades downstream CP utility. Full-size prototypes and 3D-printed optimized miniatures confirm the intended LiDAR behavior and physical producibility. \defensename shrinks this attack surface by masking locally disputed regions before trust evaluation, although residual cases show that fully neutralizing adversarial trust poisoning remains open.

In summary, our contributions are:
\begin{itemize}
    \item We identify and formalize \textit{adversarial trust poisoning}, an attribution vulnerability in which physically induced cross-view disagreement makes a consistency-based CP defense suppress a targeted benign vehicle, with no digital intrusion or fabricated V2X message.
    \item We design \attackname, a two-stage physical attack that couples physically biased shape priors with constrained, defense-aware optimization and runtime deployment policies, and \defensename, a self-reflection add-on that masks locally uncertain evidence before trust evaluation without retraining.
    \item We evaluate both on OPV2V and V2X-Real across four CP backbones and four SOTA defenses, and validate physical feasibility with real LiDAR captures of fabricated prototypes. \attackname reaches up to $88\%$ defense-level ASR and a $6.0\%$ absolute AP loss on the full benchmark (up to $91\%$ and $13\%$ points, respectively, on critical subsets if allow the adversary to select scenarios), and \defensename reduces ASR by $35$--$100\%$.
\end{itemize}

%% file: related_work.tex
\section{Related Work}
\label{sec:related-work}

\myparagraph{Collaborative Perception (CP)} CAVs pair onboard sensors, such as LiDARs and cameras, with wireless modules that let them exchange perception information over Vehicle-to-Everything (V2X) links~\cite{CAV, chen2019cooper, zhang2021emp}. CP exploits this capability to overcome single-agent limitations such as occlusion, restricted sensing range, and sparse long-range point clouds. Although CP admits multiple modalities, most representative systems build on LiDAR or LiDAR-derived bird's-eye-view (BEV) representations, which capture rich 3D semantics for object localization and cross-agent spatial alignment.

CP systems are categorized by the stage at which information is shared. \textit{Early fusion} shares raw sensor observations, typically LiDAR point clouds, which are aggregated into a common coordinate frame before detection~\cite{chen2019cooper, zhang2021emp, chen2022cooperative}. It preserves the richest geometry and helps under occlusion, but incurs high communication overhead and is sensitive to latency and pose error. \textit{Intermediate fusion} encodes each local observation into neural features, such as BEV feature maps, and communicates them for cross-agent aggregation~\cite{chen2019f, wang2020v2vnet, cui2022coopernaut, xu2022v2x, yuan2022keypoints, hu2022where2comm, zhang2023robust}. This approach trades accuracy against bandwidth, with recent systems adding graph reasoning, attention, transformers, spatial confidence maps, and pose alignment~\cite{lu2022robust,wang2025jigsawcomm}. \textit{Late fusion} shares final outputs, such as bounding boxes, confidence scores, or tracks, and merges them by object-level association~\cite{liu2019fusioneye, shi2022vips}. It costs the least bandwidth and tolerates heterogeneous detectors, but discards dense geometric and feature context, leaving performance dependent on local detection quality and cross-agent association. This taxonomy matters for security because each stage exposes a different attack surface and defense signal: raw sensor evidence, learned representations, and object-level decisions, respectively.

\myparagraph{CP Attacks}
The primary threat in CP is data fabrication, where malicious participants share crafted sensor data to perturb victim perception~\cite{tu2021adversarial, zhang2024data, wang2025threat, wang2025cp}. In early fusion, black-box ray casting reconstructs realistic yet malicious point clouds that spoof or remove objects. Offline adversarial-object generation and runtime occlusion-aware point sampling further make the fabricated data obey physical constraints and evade basic anomaly detection.

\myparagraph{CP Defenses}
To mitigate fabrication and malicious collaboration, CP defenses identify inconsistent shared data or untrusted collaborators. CAD~\cite{zhang2024data} validates cross-vehicle geometric consistency with fine-grained occupancy maps and checks final perception against the merged occupancy evidence. MATE~\cite{hallyburton2025security} targets late-fusion tracking, modeling agent and track trust as latent probabilistic states updated by pseudomeasurements derived from discrepancies between expected visibility and reported tracks. ROBOSAC~\cite{li2023among} samples collaborator subgroups until one reaches consensus. For intermediate fusion, MADE~\cite{zhao2024made} combines output-level matching with feature-level collaborative reconstruction statistics calibrated on benign data in a semi-supervised multi-test framework, while LUCIA~\cite{wang2025threat} computes feature-consistency trust scores that modulate each collaborator's attention contribution. We evaluate \attackname against CAD, MATE, MADE, and LUCIA, covering occupancy consistency, late-fusion trust estimation, feature-reconstruction anomaly detection, and attention-level trust modulation. We analyze ROBOSAC separately because its scene-level consensus responds differently to localized single-object discrepancies.

\myparagraph{Physical Adversarial Object Attacks}
Physical adversarial attacks deceive sensors by modifying the environment rather than shared data. Optimizing the 3D geometry of simple or everyday objects, such as traffic cones or vehicle-mounted luggage, induces misclassification or detection failure in LiDAR-based systems~\cite{cao2019adversarial, cao2021invisible, tu2020physically, zhu2024ae}. Multimodal designs disrupt visual and geometric features simultaneously, hiding the object from both camera and LiDAR detectors~\cite{cao2021invisible}. Object-agnostic variants instead exploit spatial vulnerabilities in LiDAR detectors by corrupting the geometric features of the surrounding context, allowing arbitrary physical shapes to serve as adversarial carriers~\cite{zhu2021can}.

These attacks target one vehicle's perception output or seek a failure consistent across viewpoints. \attackname reuses established differentiable LiDAR rendering and detector relaxations but optimizes a different security objective: viewpoint-dependent perception is the mechanism, while the downstream defense's attribution decision is the target. The consequence therefore extends beyond a detection error, as the defense can suppress a benign sensing contributor with no compromised vehicle and no fabricated V2X message.

%% file: threat_model.tex
\section{Threat Model}
\label{sec:threat-model}

\myparagraph{System assumptions}
We consider a \textbf{LiDAR-based CP system} operating under \textbf{intermediate} and \textbf{late fusion}. Each vehicle processes its raw sensor observations locally and shares either compressed neural feature maps for intermediate fusion or final perception outputs, such as 3D bounding boxes and occupancy maps, for late fusion over V2X communication. The ego vehicle aggregates these messages into a multi-agent scene representation. We exclude early fusion because transmitting raw LiDAR point clouds incurs high communication overhead~\cite{chen2019f}, and few existing defenses support this setting.

To mitigate data fabrication, the system employs \textit{consistency-based defenses} that validate cross-agent agreement and downweight or exclude out-of-consensus contributions. We consider four primary approaches. CAD~\cite{zhang2024data} enforces occupancy-map consistency. MATE~\cite{hallyburton2025security} performs late-fusion Bayesian trust estimation. MADE~\cite{zhao2024made} detects anomalies at the intermediate-fusion stage. LUCIA~\cite{wang2025threat} assigns trust scores from feature-level consistency during intermediate fusion. We analyze ROBOSAC~\cite{li2023among} separately to contrast these localized checks with scene-level consensus, which may absorb a single-object discrepancy without changing the subgroup decision.

\myparagraph{Attack goal}
The adversary's objective is \textbf{targeted trust poisoning} against a designated benign CAV. We define two collaborator roles:
\begin{itemize}
    \item \textbf{Victim}: the benign collaborator whose viewpoint is intentionally made vulnerable and whose shared perception should later be judged suspicious.
    \item \textbf{Non-victims}: benign collaborators that observe the same adversarial object from non-vulnerable angles and provide the apparent consensus evidence.
\end{itemize}
The attack succeeds when the defense treats the victim as the source of the disagreement and consequently downweights, excludes, or labels it as malicious. The evidence that triggers this decision depends on the defense interface and may be an occupancy conflict, a detection or track disagreement, an intermediate-feature inconsistency, or an anomaly-score deviation. The perception discrepancy is thus the physical mechanism, false attribution is the security outcome, and the resulting loss of a benign sensing contribution can degrade collaborative perception.

\myparagraph{Capabilities}
The adversary introduces a \textit{physical adversarial object}, an attacker-controlled 3D structure whose geometry and pose induce viewpoint-dependent LiDAR returns. The attacker can control the object's shape, position, and orientation and deploy it in realistic ways, such as mounting it on an attacker-controlled vehicle, towing it as an attached structure, or placing it at feasible roadside locations such as curbs, medians, or construction zones. The adversary may further select scenarios and victims in which the relative CAV geometry creates asymmetric visibility across agents. Importantly, the attacker does not compromise the victim, inject digital data, or interfere with V2V communication.

We focus on physical objects rather than active sensor attacks such as LiDAR spoofing~\cite{jin2023pla} or jamming~\cite{petit2014potential}. Adversarial physical objects are passive and persistent. They require no signal-injection hardware, synchronization with the target sensor, or online control. Moreover, they generate contiguous, physically plausible LiDAR evidence over long temporal horizons, such as while remaining in the victim's lane, producing consistent signals that can influence temporal defenses.

\myparagraph{Knowledge}
The attack requires two phases of knowledge. During offline preparation, the attacker has \emph{white-box} knowledge of the target perception model, defense logic, and decision thresholds. This is the standard assumption in prior physical adversarial attacks on collaborative and single-vehicle LiDAR perception~\cite{tu2021adversarial,cao2019adversarial,cao2021invisible,tu2020physically,zhu2024ae}. At runtime, the attacker requires only \emph{environmental} knowledge. The attacker observes the road layout and approximate CAV positions, then selects a victim and object pose that realize the vulnerable and non-vulnerable viewing angles. No runtime access to internal model states, V2X messages, or gradients is required.

%% file: design.tex
\section{Design of \attackname}
\label{sec:design}

\subsection{Problem Formulation}
\label{subsec:problem-formulation}

We formulate \attackname as a view-conditioned physical adversarial object optimization problem in CP. Its immediate objective is to make benign vehicles perceive the same physical object differently according to their viewpoints. We evaluate this behavior at two levels. The perception-level objective measures whether the object produces the intended cross-view discrepancy, while the defense-level objective measures whether that discrepancy causes a consistency-based defense to penalize a designated benign collaborator.

Let $S$ denote the physical adversarial object. We partition its viewing-angle space into two disjoint sets:
\[
\Theta = \Theta_v \cup \Theta_n,
\]
where $\Theta_v$ is the \emph{vulnerable angle set} corresponding to the victim view, and $\Theta_n$ is the \emph{non-vulnerable angle set} corresponding to non-victim collaborators' views.

Let $f(\cdot)$ denote the LiDAR-based detector used by each agent, and define
\[
D(\theta; S) =
\begin{cases}
1, & \text{if } f \text{ detects } S \text{ from angle } \theta, \\
0, & \text{otherwise}.
\end{cases}
\]
For detector-based perception, the desired view-conditioned discrepancy is
\[
D(\theta; S) =
\begin{cases}
0, & \forall \theta \in \Theta_v, \\
1, & \forall \theta \in \Theta_n.
\end{cases}
\]
The object therefore disappears from victim angles while remaining visible from non-victim angles under the same perception model. This condition defines the perception-level objective used to measure whether \attackname creates the intended physical discrepancy.

We next formulate the downstream defense objective. Let $g(\cdot)$ denote a CP defense executed by a non-victim ego vehicle, and let
\[
\Phi_g(i \mid \theta; S)
\]
be the defense-specific suspicion score assigned to collaborator $i$ when the ego observes the scene from a non-victim view $\theta\in\Theta_n$. A larger $\Phi_g$ indicates greater suspicion. The defense-level objective is to make the victim $v$ appear more suspicious than every non-victim collaborator $n$:
\[
\Phi_g(v \mid \theta; S) > \Phi_g(n \mid \theta; S),
\qquad
\forall \theta \in \Theta_n.
\]
For threshold-based defenses, the victim should cross the anomaly boundary while non-victims remain below it:
\[
\Phi_g(v \mid \theta; S) \ge \tau_g,
\qquad
\Phi_g(n \mid \theta; S) < \tau_g,
\qquad
\forall \theta \in \Theta_n,
\]
where $\tau_g$ is the decision threshold of defense $g$. The evidence consumed by $g$ may be a detection, occupancy, track, or intermediate-feature inconsistency. Accordingly, \attackname either optimizes the perception discrepancy or directly optimizes a defense-specific score. The former establishes the physical attack mechanism, and the latter tests whether the same mechanism can manipulate a deployed defense.

\subsection{Design Challenges and Insights}
\label{subsec:design-challenges}

The formulation defines the desired end state but not how a physical object realizes it. We organize the design of \attackname around two challenges and their insights, deferring concrete instantiations to Section~\ref{subsec:offline-optimization}.

\begin{figure}[t]
    \centering
    \begin{subfigure}[t]{0.32\columnwidth}
        \centering
        \includegraphics[width=\columnwidth]{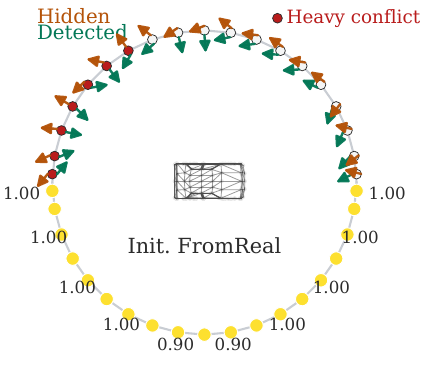}
        \caption{\textsc{FromReal} prior}
        \label{subfig:real-prior}
    \end{subfigure}
    \hfill
    \begin{subfigure}[t]{0.32\columnwidth}
        \centering
        \includegraphics[width=\columnwidth]{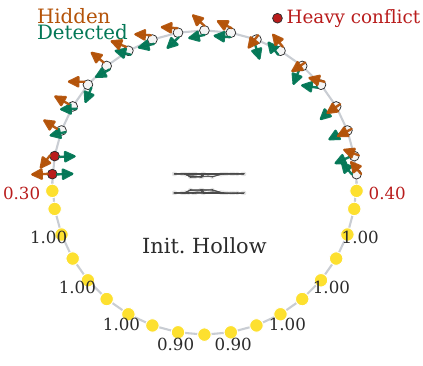}
        \caption{Hollow prior}
        \label{subfig:hollow-prior}
    \end{subfigure}
    \hfill
    \begin{subfigure}[t]{0.32\columnwidth}
        \centering
        \includegraphics[width=\columnwidth]{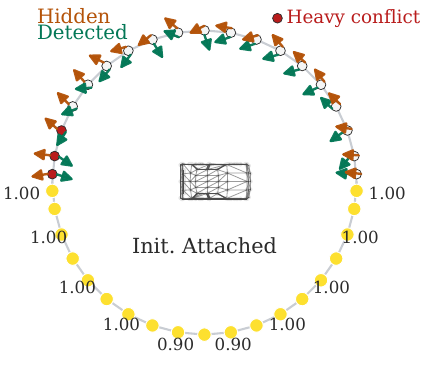}
        \caption{Attached prior}
        \label{subfig:attached-prior}
    \end{subfigure}
    \caption{Gradient-conflict diagnostics for three shape priors. In each polar plot, the upper semicircle shows the conflict between victim-disappearance and non-victim-detection gradients across viewing angles, while the lower semicircle shows detector confidence over the same angles. The \textsc{FromReal} prior exposes strong conflict on shared surfaces. \textsc{Hollow} and \textsc{Attached} reduce it by physically biasing one side of the view-conditioned objective.}
    \label{fig:gradient-conflict}
\end{figure}

\challenge{Conflicting view-conditioned objectives}\label{ch:conflict}
The same 3D object must disappear from victim views and remain detectable from non-victim views, yet the two sets of views often fall on overlapping surfaces, so moving a shared vertex to suppress victim-view detection can undermine the evidence that preserves non-victim detection.

Let $\mathbf{V}_i\in\mathbb{R}^3$ be the coordinates of mesh vertex $i$, and let $c_\theta\in[0,1]$ be the detector confidence for the adversarial object from viewing angle $\theta$. We use $\ell_{\mathrm{hide}}(c)=-\log(1-c)$ to suppress confidence and $\ell_{\mathrm{show}}(c)=-\log(c)$ to preserve it, giving expected victim- and non-victim-view gradients at vertex $i$ of
\[
\begin{aligned}
g_v^i & = \mathbb{E}_{\theta\in\Theta_v}\!\left[\nabla_{\mathbf{V}_i}\ell_{\mathrm{hide}}(c_\theta)\right], \\
g_n^i & = \mathbb{E}_{\theta\in\Theta_n}\!\left[\nabla_{\mathbf{V}_i}\ell_{\mathrm{show}}(c_\theta)\right],
\end{aligned}
\]
where each expectation averages over its corresponding viewing-angle set. The two objectives conflict on vertex $i$ when $\cos(g_v^i,g_n^i)<0$: the updates oppose each other and progress toward one or both view-conditioned goals stalls. Figure~\ref{subfig:real-prior} visualizes this conflict on a closed vehicle prior.

\insight{Use a shape initialization that physically induces view-conditioned perception discrepancy.}\label{ins:shape-prior}
A well-chosen initial geometry already satisfies or preserves one side of the view-conditioned objective, so the optimizer need not drive opposing losses over the same surfaces: a hollow or thin-board structure lets LiDAR rays pass from vulnerable angles while presenting an occupied profile from non-vulnerable ones, and a closed vehicle prior preserves vehicle-like side and frontal returns while still allowing localized changes to the rear silhouette. Figures~\ref{subfig:hollow-prior} and~\ref{subfig:attached-prior} show the resulting reduction in gradient conflict.

\insight{Object optimization is a multi-parameter trade-off among effectiveness, robustness, and stealthiness.}\label{ins:tradeoff}
Four parameters govern this trade-off. A narrow vulnerable angle $\Theta_v$ paired with a wide non-vulnerable angle $\Theta_n$ maximizes perception asymmetry, but $\Theta_v$ cannot be too narrow without losing reliability against victim-pose error, and widening the gap amplifies the gradient conflict of Challenge~\ref{ch:conflict} at training time. Likewise, a larger perturbable area or per-vertex perturbation bound improves attack effectiveness, yet produces shapes that are visibly anomalous to non-victims and that overfit the differentiable surrogate, hurting robustness on the deployed perception pipeline.

\attackname therefore couples a \textit{physically biased shape initialization}, which resolves one side of the conflicting objective, with \textit{constrained optimization} through vertex masks, per-vertex displacement bounds, and conservative angle ranges, which refines the other side without destroying useful geometry or producing unrealistic shapes.

\challenge{Realistic deployment beyond view discrepancy}\label{ch:deploy}
A view-conditioned discrepancy is not sufficient on its own. The attack must also (i) flow through the specific defense's decision logic so the inconsistency is attributed to the victim, (ii) yield a physically realizable shape that survives fabrication and real LiDAR sampling, and (iii) be deployed where excluding the victim actually degrades CP utility.

\insight{Match the optimization objective to the defense interface.}\label{ins:differentiable-defense}
CAD and MATE consume detection-derived late-fusion outputs, so \attackname optimizes the view-conditioned perception discrepancy without differentiating through either defense: the resulting occupancy-map discrepancy activates CAD, and the bounding-box and track discrepancy activates MATE. LUCIA and MADE instead operate on intermediate features or reconstruction statistics, so \attackname differentiates their relevant scoring branches.

\insight{Mesh complexity, per-vertex perturbation bounds, and smoothness regularization jointly control physical realizability.}\label{ins:realizability}
The vertex count of the initial mesh sets the spatial frequency of expressible perturbations, per-vertex displacement bounds prevent extreme deformation, and Laplacian smoothing penalizes high-frequency artifacts that arise only in the differentiable surrogate, together keeping the optimized geometry fabricable and stable under real LiDAR sampling.

\begin{figure}[t]
    \centering
    \begin{subfigure}[t]{0.48\columnwidth}
        \centering
        \includegraphics[width=\columnwidth]{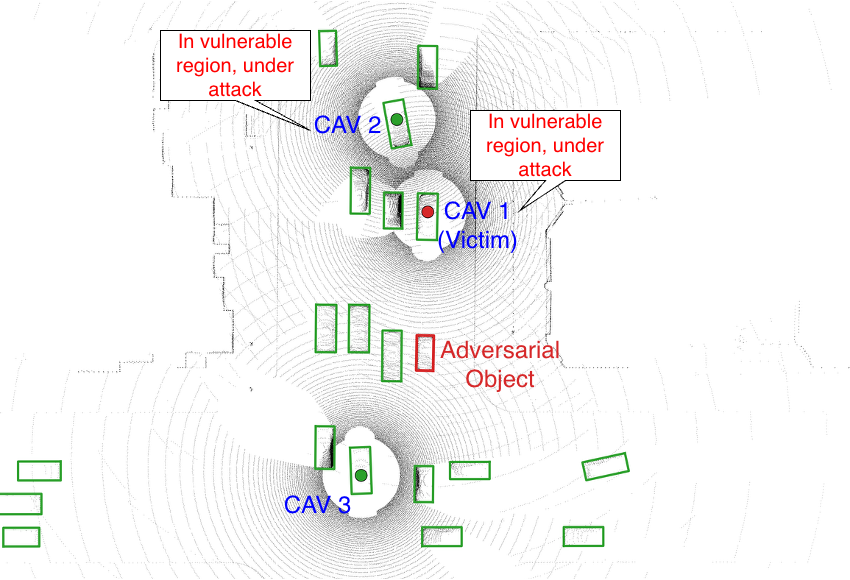}
        \caption{Hard: non-victim close to victim}
    \end{subfigure}
    \hfill
    \begin{subfigure}[t]{0.48\columnwidth}
        \centering
        \includegraphics[width=\columnwidth]{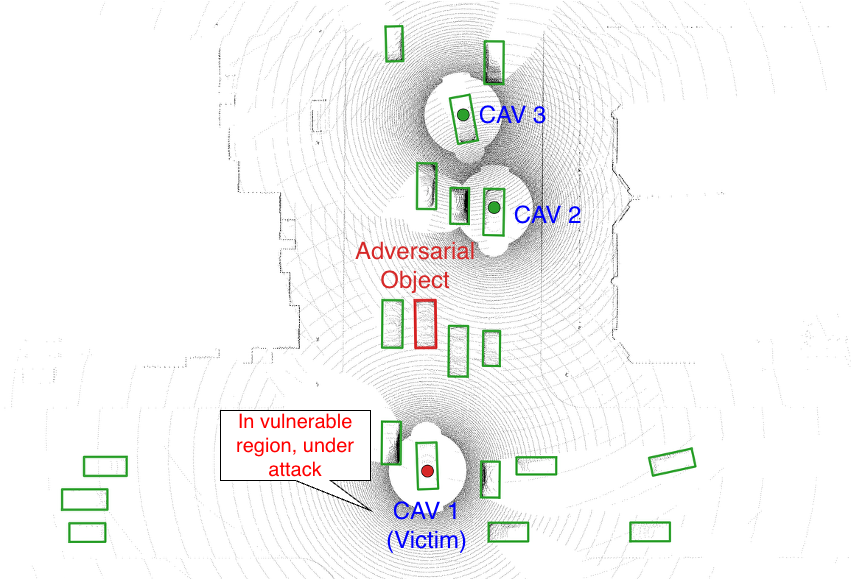}
        \caption{Easy: non-victim with distinct view angle}
    \end{subfigure}\\[2pt]
    \begin{subfigure}[t]{0.48\columnwidth}
        \centering
        \includegraphics[width=\columnwidth]{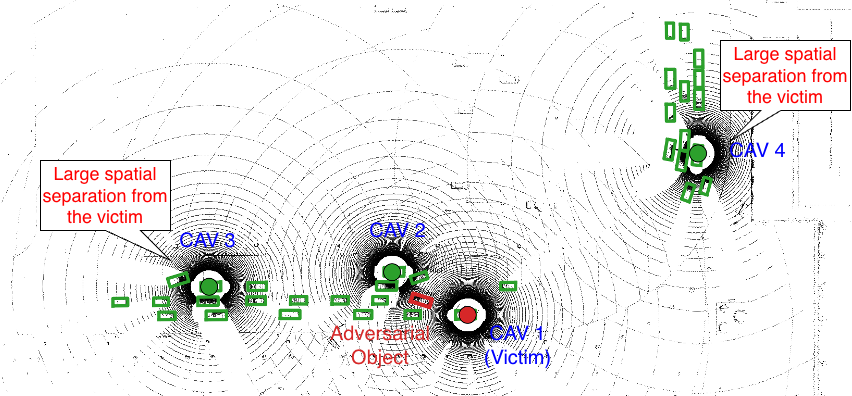}
        \caption{High safety impact: victim's view is uniquely informative}
    \end{subfigure}
    \hfill
    \begin{subfigure}[t]{0.48\columnwidth}
        \centering
        \includegraphics[width=\columnwidth]{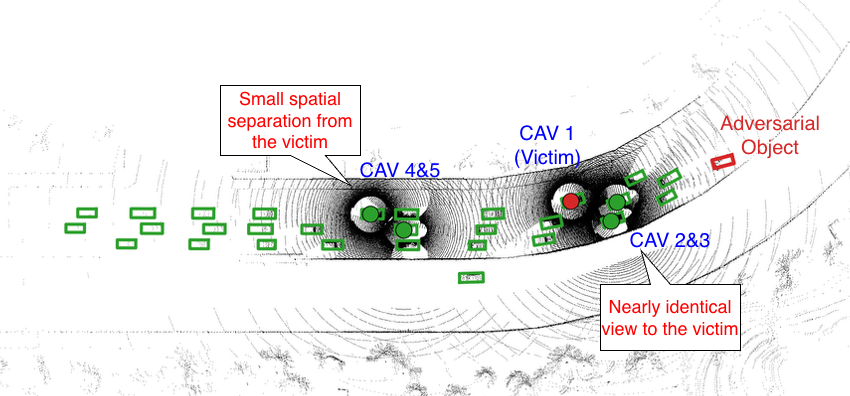}
        \caption{Low safety impact: non-victims cover the same regions}
    \end{subfigure}
    \caption{Illustration of Insight~\ref{ins:scenario}. The attack's success rate and safety impact are sensitive to the on-road scenarios.}
    \label{fig:scenario-sensitivity}
\end{figure}

\insight{Select scenarios where CP performance is sensitive to the victim's contribution.}\label{ins:scenario}
Scene geometry governs both attack feasibility and safety impact (Figure~\ref{fig:scenario-sensitivity}). Relative CAV positions and host-vehicle silhouettes determine whether the vulnerable angle $\Theta_v$ is reliably realized: a non-victim close to the victim shares a similar viewing angle and perceives more consistently, making the attack harder, whereas a distinct non-victim view makes asymmetric LiDAR returns easier to induce. Separately, the victim's unique sensing coverage (e.g., regions only it can observe) determines how much downstream perception is lost when the victim is excluded. The attacker therefore assesses candidate scenarios, picks road segments and timings where collaboration is most informative, and aligns victim/object geometry to those moments.

\subsection{Attack Overview}
\label{subsec:attack-overview}

\begin{figure}[t]
    \centering
    \includegraphics[width=\columnwidth]{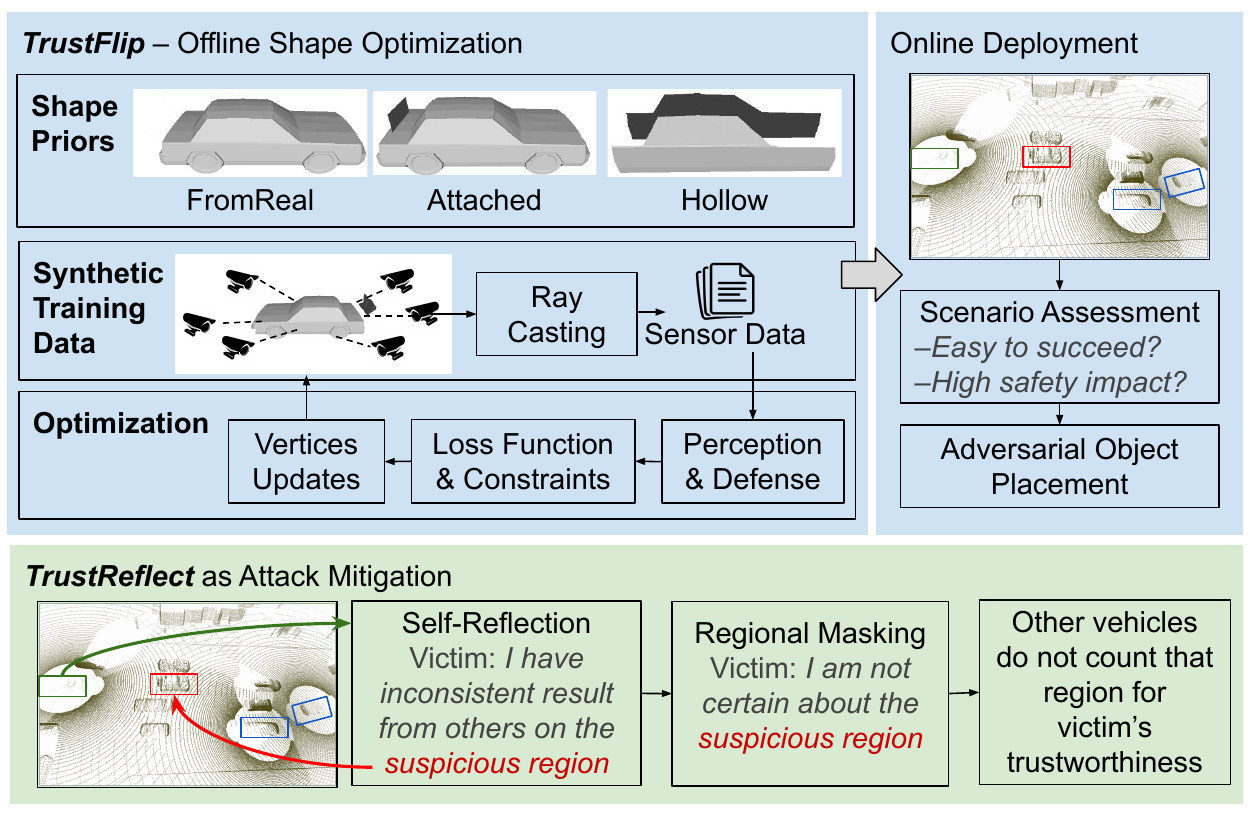}
    \caption{Overview of \attackname and \defensename.}
    \label{fig:methodology}
\end{figure}

Following Section~\ref{subsec:design-challenges}, the attack proceeds in two stages (Figure~\ref{fig:methodology}). The \emph{offline stage} (Section~\ref{subsec:offline-optimization}) solves a constrained mesh-optimization problem that produces a physical adversarial object whose geometry simultaneously realizes the shape prior of Insight~\ref{ins:shape-prior}, the detector- or defense-level objective of Insight~\ref{ins:differentiable-defense}, and the realizability constraints of Insight~\ref{ins:realizability}. The \emph{online stage} (Section~\ref{subsec:online-deployment}) selects a scenario, a victim, and an object pose that realize the trained vulnerable and non-vulnerable view sets at runtime, following Insight~\ref{ins:scenario}. The runtime attack is purely physical and requires no digital intrusion. As mitigation, Section~\ref{subsec:mitigation} introduces \defensename, a lightweight and defense-compatible self-reflection layer that reuses each deployed defense's own mechanism and adds only a trust-mask field to V2X messages, allowing the same integration pattern to layer on CAD, MATE, LUCIA, and MADE without retraining.

\subsection{Offline Shape Optimization}
\label{subsec:offline-optimization}

\myparagraph{Shape priors}
Following Insight~\ref{ins:shape-prior}, we instantiate three shape priors with different physical biases (Figure~\ref{fig:methodology}):
\begin{itemize}
    \item \textbf{\textsc{FromReal}}: a normal car mesh whose deformation is confined to a vertex mask. Non-victim views retain the benign vehicle prior, while the rear surface can be perturbed to suppress victim-view detection.
    \item \textbf{\textsc{Attached}}: a small planar occluder mounted on the trunk of a normal host vehicle. Non-victims still observe the unmodified host vehicle, while the rear silhouette is locally distorted by the occluder.
    \item \textbf{\textsc{Hollow}}: a standalone structure of two parallel thin boards with a hollow center. Rays from $\Theta_v$ pass through the gap and hit the ground, while rays from $\Theta_n$ strike the side boards and form an occupied car-like profile.
\end{itemize}
Among the three, \textsc{Hollow} is the only prior that produces a free-versus-occupied cross-agent conflict, which is required by occupancy-map defenses such as CAD~\cite{zhang2024data}. Closed priors keep the victim's occupancy map occupied.

The three priors also differ in \textit{stealthiness}. \textsc{Attached} is the most stealthy: the adversarial component is small, largely occluded by the host vehicle from non-victim views, and still plausible from the victim view because objects attached to a trunk or carried behind a vehicle are common in traffic. \textsc{FromReal} is less stealthy because its own body is perturbed, but non-victims still observe a partial or complete car-shaped object. \textsc{Hollow} is the least stealthy: although it creates the strongest free-versus-occupied conflict, its shape is distinctive and easier to notice outside the perception pipeline.

\myparagraph{Optimization objectives}
For detector-level optimization, \attackname uses a view-conditioned confidence loss with $\ell_{\mathrm{hide}}(c)=-\log(1-c)$ on victim views and $\ell_{\mathrm{show}}(c)=-\log(c)$ on non-victim views, combined as
\[
\mathcal{L}_{\mathrm{vis}} = \alpha\,\mathbb{E}_{\theta\in\Theta_v}\ell_{\mathrm{hide}}(c_\theta) + \beta\,\mathbb{E}_{\theta\in\Theta_n}\ell_{\mathrm{show}}(c_\theta),
\]
where the weights $(\alpha,\beta)$ depend on which side of the objective the shape prior already satisfies, as detailed below.

Following Insight~\ref{ins:differentiable-defense}, the adversarial loss for LUCIA and MADE instead optimizes a defense-specific suspicion margin from non-victim ego views.

\textbf{LUCIA~\cite{wang2025threat}.} The post-softmax trust score normalizes away gradient magnitude, so \attackname optimizes the raw pre-softmax inconsistency on the target-local feature region. Let $s_v^{(\theta)}$ and $s_n^{(\theta)}$ denote the inconsistency scores assigned by LUCIA to the victim and to a peer non-victim collaborator under non-victim ego view $\theta$:
\[
\ell_{\mathrm{LUCIA}}=\frac{1}{|\Theta_n|}\sum_{\theta\in\Theta_n}\!\bigl(w_n\,s_n^{(\theta)}-w_v\,s_v^{(\theta)}\bigr).
\]

\textbf{MADE~\cite{zhao2024made}.} Let $\hat{s}_k$ be MADE's normalized suspicion score for inspected collaborator $k$ (instantiated by its reconstruction score, or the weighted match+reconstruction combination). With $\mathcal{V}$ and $\mathcal{N}$ denoting the inspection sets in which the victim and a non-victim peer is inspected, respectively:
\[
\ell_{\mathrm{MADE}}=-\frac{1}{|\mathcal{V}|}\!\sum_{k\in\mathcal{V}}\hat{s}_k + \beta\frac{1}{|\mathcal{N}|}\!\sum_{k\in\mathcal{N}}\hat{s}_k.
\]

\myparagraph{Shape-prior-specific weight setting}
The visibility loss above is shared across visibility-driven defenses, but the loss-weight choice $(\alpha,\beta)$ depends on which side of the conflict the shape prior already resolves (Insight~\ref{ins:shape-prior}):
\begin{itemize}
    \item \textbf{\textsc{FromReal}}: $(\alpha,\beta)=(1,0)$ on a vertex mask. The closed car prior already supports non-victim detection, so only the victim-disappearance loss is active.
    \item \textbf{\textsc{Attached}}: $(\alpha,\beta)=(1,0)$. The host vehicle's body provides the non-victim returns, and the optimizer concentrates on the planar occluder that suppresses the rear silhouette.
    \item \textbf{\textsc{Hollow}}: $(\alpha,\beta)=(0,1)$. The hollow geometry already eliminates victim-view returns, so the optimizer reinforces non-victim visibility on the side boards.
\end{itemize}

\myparagraph{Constraints and smoothness}
Per Insight~\ref{ins:realizability}, each shape prior carries a constraint set $\mathcal{C}$ that confines optimization to a fabricable, sensor-stable region:
\begin{itemize}
    \item \textbf{\textsc{FromReal}}: optimization is restricted to a vertex mask $\mathcal{M}$, with a global size bound $\mathbf{b}_s$ on the mesh extent.
    \item \textbf{\textsc{Attached}}: a translation bound $\mathbf{b}_t$ on the mounted occluder, plus a size bound $\mathbf{b}_s$ on its mesh extent.
    \item \textbf{\textsc{Hollow}}: a per-vertex displacement bound $\mathbf{b}_v$ on $\mathbf{V}-\mathbf{V}_0$, preserving the parallel-board pattern of the initialization.
\end{itemize}
On top of $\mathcal{C}$ we add a Laplacian smoothness penalty $\mathcal{L}_{\mathrm{lap}}=\sum_i\|\delta_i\|_2^2$, where $\delta_i$ is the offset from each vertex to its one-ring centroid, to suppress high-frequency artifacts that exist only in the differentiable surrogate.

\myparagraph{Differentiable surrogates}
LiDAR detectors contain non-differentiable preprocessing. During training, \attackname replaces hard BEV grid or pillar discretizations with smooth point-to-cell assignment and approximates other non-differentiable components. Surrogates are used only for gradient computation. The final adversarial object is evaluated against the original perception and defense pipelines (Section~\ref{subsec:experiment-setup}).

\myparagraph{Optimization views}
Offline optimization samples representative views $\mathcal{S}_v\subset\Theta_v$ and $\mathcal{S}_n\subset\Theta_n$ rather than full traffic scenes. These views isolate the adversarial object with local ground support so that gradients focus on object geometry and view-conditioned discrepancy, while final evaluation is performed in complete CP scenes with the original perception and defense pipelines.

\begin{algorithm}[t]
\small
\caption{\attackname offline mesh optimization.}
\label{alg:adv-mesh-optimization}
\begin{algorithmic}[1]
\Statex \textbf{Input:} initial mesh $(\mathbf{V}_0,\mathbf{F})$; views $\mathcal{S}_v,\mathcal{S}_n$; task loss $\mathcal{L}_{\mathrm{task}}\!\in\!\{\mathcal{L}_{\mathrm{vis}},\ell_{\mathrm{LUCIA}},\ell_{\mathrm{MADE}}\}$; shape constraint $\mathcal{C}$; smoothness weight $\lambda$; learning rate $\eta$; epochs $E$.
\Statex \textbf{Output:} optimized mesh $(\mathbf{V},\mathbf{F})$.
\State $\mathbf{V}\leftarrow\mathbf{V}_0$
\For{epoch $=1,\dots,E$}
    \State $\mathcal{L}\leftarrow \mathcal{L}_{\mathrm{task}}(\mathbf{V},\mathbf{F},\mathcal{S}_v,\mathcal{S}_n) + \lambda\,\mathrm{LaplacianSmooth}(\mathbf{V},\mathbf{F})$
    \State $\mathbf{V}\leftarrow \mathbf{V} - \eta\,\nabla_{\mathbf{V}}\mathcal{L}$ \Comment{gradient descent step}
    \State $\mathbf{V}\leftarrow \mathrm{Proj}_{\mathcal{C}}(\mathbf{V})$ \Comment{shape-constraint projection}
\EndFor
\State \Return $(\mathbf{V},\mathbf{F})$
\end{algorithmic}
\end{algorithm}

\myparagraph{Parameter selection}
The vulnerable angle $\Theta_v$ is set to a narrow cone (e.g., $\ang{5}$) directly behind the object, modeling a realistic same-lane towing geometry. $\Theta_n$ is its complement on the surrounding hemisphere. Insight~\ref{ins:tradeoff} captures the operative trade-offs: a narrower $\Theta_v$ paired with a wider $\Theta_n$ improves perception asymmetry but reduces robustness to victim-pose error and amplifies the gradient conflict, while larger displacement bounds increase ASR but hurt non-victim plausibility. We use a single set of defaults for $\Theta_v$, $\Theta_n$, the perturbed area, the displacement bound, and $\lambda$ (Section~\ref{subsec:experiment-setup}). Their sensitivity is studied in Section~\ref{subsubsec:attack-parameters}.

\subsection{Online Deployment}
\label{subsec:online-deployment}

\begin{algorithm}[t]
\small
\caption{\attackname online deployment.}
\label{alg:online-deployment}
\begin{algorithmic}[1]
\Statex \textbf{Input:} candidate scenarios $\Sigma$ (road segments + time windows), public HD map, observable CAV poses $\{\mathbf{x}_c\}$, optimized mesh $S$.
\Statex \textbf{Output:} chosen scenario $\sigma^\ast$, victim $v^\ast$, object pose $p^\ast$.
\State \textbf{Scenario assessment:}
\Statex \quad $\sigma^\ast \leftarrow \arg\max_{\sigma\in\Sigma}\;\bigl[\omega_o\,O(\sigma) + \omega_c\,|\mathcal{C}(\sigma)|\bigr]$
\Statex \quad where $O(\sigma)$ is the occlusion density and $|\mathcal{C}(\sigma)|$ is the CAV count.
\State \textbf{Object pose:} pick towable pose $p^\ast$ in $\sigma^\ast$ aligning the rear cone of $S$ with the dominant traffic heading.
\State \textbf{Victim selection:}
\Statex \quad $v^\ast \leftarrow \arg\max_{c\in\mathcal{C}(\sigma^\ast)}\;\bigl[\mathbb{1}\{\angle(c,p^\ast)\in\Theta_v\}\cdot R(c)\bigr]$
\Statex \quad subject to $\exists\, c'\neq v^\ast:\;\angle(c',p^\ast)\in\Theta_n$,
\Statex \quad where $R(c)$ is $c$'s estimated unique sensing coverage (pixels visible only to $c$).
\State \Return $(\sigma^\ast, v^\ast, p^\ast)$
\end{algorithmic}
\end{algorithm}

Algorithm~\ref{alg:online-deployment} uses only public map data and observable CAV poses. It does not intercept V2X messages.

\myparagraph{Deployment policy}
(1) \emph{Scenario assessment}: The score $\omega_o O+\omega_c|\mathcal{C}|$ favors scenes with heavier occlusion, where the victim's observations are harder to corroborate, and more CAVs, which provide more cross-agent trust evidence. Both factors are available from public maps and observable traffic. (2) \emph{Victim selection}: The angle constraint places the victim in $\Theta_v$, while $R(c)$ favors a CAV with uniquely informative sensing coverage so that suppressing its contribution has greater utility impact. The non-victim constraint requires corroborating evidence from at least one CAV in $\Theta_n$. (3) \emph{Object placement and trajectory control}: For \textsc{FromReal} and \textsc{Hollow}, the attacker tows the object behind an attacker-driven vehicle in the victim's lane and maintains a longitudinal gap consistent with $\Theta_v$. For \textsc{Attached}, the structure is mounted behind an attacker-driven host vehicle, whose body supplies benign returns from adjacent non-victim views. The attacker adjusts speed and lane position frame by frame to maintain the required angles. Occasional violations do not invalidate the attack because consistency-based defenses react to repeated disagreement, and a moderate fraction of valid frames can lower trust within a temporal window such as MATE's 10-frame window (Section~\ref{subsubsec:mate-effectiveness}).

\myparagraph{Robustness}
Offline training samples variations in distance, heading, ground-plane reflectivity, and partial occlusion, and the same optimized mesh is reused across scenes without retuning. Deployment then requires maintaining the relative viewing geometry. Table~\ref{tab:deployment-robustness} evaluates tolerance to imprecise object heading, a key source of real-world deployment error.

\subsection{Mitigation: \defensename}
\label{subsec:mitigation}

The key observation behind \defensename is that the cross-view inconsistency \attackname creates is also visible from the \emph{victim's own} sensor. If the victim runs the same consistency-based defense locally, the defense will flag a malicious-collaborator-like pattern in the very region the adversarial object occupies. \defensename adds a single \emph{self-reflection} edge to the V2X protocol (Figure~\ref{fig:methodology}): each benign vehicle re-runs the deployed defense's scoring rule against its own perception output and, if the local score crosses the malicious-collaborator threshold, treats that region as suspicious and uncertain. The vehicle then (i) localizes the suspicious region using the defense's own prediction- or feature-level output, (ii) attaches a per-frame \emph{trust mask} declaring that region uncertain to its shared data, and (iii) downstream consumers exclude the masked region from their trust scores or fusion weights. The reflection can leverage the existing defenses without training new models.

\emph{A malicious vehicle cannot exploit \defensename for data fabrication.} Although the attacker may mark the attacked region as uncertain to avoid trust-score degradation, the fabricated data from that region is also excluded from fusion and therefore has no effect.

Region selection and masking are defense-specific:
\begin{itemize}
    \item \textbf{CAD~\cite{zhang2024data}}: the reflection compares the ego's occupancy map with a single-vehicle baseline. Cells where the ego's own occupancy disagrees with the baseline form the trust mask, and CAD's cross-agent occupancy check ignores those cells.
    \item \textbf{MATE~\cite{hallyburton2025security}}: the reflection compares its ego-view predicted tracks with aggregated tracks. Tracks for which the ego's reported visibility is inconsistent with the aggregator are excluded from the pseudomeasurement update.
    \item \textbf{LUCIA~\cite{wang2025threat}}: the reflection computes the ego's intermediate feature consistency against an ego-only feature reconstruction. Inconsistent BEV cells form the mask, and LUCIA recomputes its trust score on the unmasked feature region.
    \item \textbf{MADE~\cite{zhao2024made}}: the reflection runs MADE's reconstruction branch on the ego's own feature. Cells with high ego-side reconstruction error are added to the mask, and the following reconstruction losses are computed on the unmasked subset.
\end{itemize}
In each case, the mask emerges from the same scoring machinery the defense already uses on peers, so \defensename layers on top of the existing defenses without retraining.

%% file: evaluation.tex
\section{Evaluation}
\label{sec:evaluation}

\subsection{Experiment Setup}
\label{subsec:experiment-setup}

\myparagraph{Data collection}
We evaluate \attackname on two CP datasets, OPV2V~\cite{xu2022opv2v} and V2X-Real~\cite{xiang2024v2x}, plus two physical LiDAR studies detailed in Section~\ref{subsubsec:physical-experiment}:

\noindent(1) \emph{OPV2V:} OPV2V provides 430 10-frame scenarios with 2--5 CAVs. We randomly designate one victim and replace a non-collaborating vehicle with an adversarial object within 40~m, rear-facing and visible to the victim and at least one non-victim. The \emph{base} set uses all scenarios, while \emph{critical} subsets apply the scenario and victim selection of Section~\ref{subsec:online-deployment}.

\noindent(2) \emph{V2X-Real}: A large-scale real-world CP benchmark with five collaborating agents. We reconstruct adversarial-object returns within its recorded LiDAR scenes. Compared with OPV2V, V2X-Real contains more sensor noise and calibration variation, which also degrade benign CP predictions.

\noindent(3) \emph{Real-world full-size captures}: On a closed, unoccupied road, with all vehicles and prototypes stationary, a Velodyne VLP-32C captures full-size \textsc{Hollow} and \textsc{Attached} prototypes from rear vulnerable views in $\Theta_v$ and side non-vulnerable views in $\Theta_n$.

\noindent(4) \emph{Real-world miniature captures}: A Unitree 4D L2 LiDAR captures 1:35-scale 3D prints of the initial and optimized versions of all three shape priors.

\myparagraph{Models and defenses}
We evaluate late-fusion PIXOR~\cite{yang2018pixor} and PointPillars~\cite{lang2019pointpillars}, and intermediate-fusion Attentive Fusion~\cite{xu2022opv2v} and Where2comm~\cite{hu2022where2comm,xu2022opv2v}, both with a PointPillars backbone. The defenses are CAD~\cite{zhang2024data}, MATE~\cite{hallyburton2025security}, LUCIA~\cite{wang2025threat}, and MADE~\cite{zhao2024made}. MATE uses its 10-frame, 5-track default, LUCIA uses CR 1, and MADE follows its semi-supervised protocol. We analyze ROBOSAC~\cite{li2023among} separately in Section~\ref{subsubsec:robosac-effectiveness}.

\myparagraph{Implementation}
We implement a differentiable LiDAR renderer based on M\"oller--Trumbore ray tracing~\cite{moller2005fast} and prior differentiable relaxations~\cite{liu2023slowlidar,cao2021invisible,jang2016categorical}. Optimization uses 300 Adam epochs at learning rate 0.01, a $\ang{5}$ vulnerable cone, non-vulnerable views at least $\ang{20}$ away, $\alpha=\beta=1$, and $\lambda=0.001$. \textsc{Hollow} bounds vertex deformation to 0.1~m, while \textsc{Attached} bounds translation and mesh size. LiDAR simulation follows OPV2V's 64-channel, 70~m specification. We test frame-specific heading error and two 50-scenario critical subsets favoring either cross-view inconsistency or unique victim contribution. All main OPV2V results use the full benchmark. Appendix~\ref{app:implementation-details} provides details.

\myparagraph{Evaluation metrics}
We report view-conditioned target detection, defense-level ASR, and AP to measure perception discrepancy, victim penalization, and downstream utility, respectively. Rates are computed over valid frames or 10-frame trajectories as specified below.

\emph{View-conditioned target detection}: for an ego view, let $\mathrm{IoU}^*$ be the maximum IoU between the ground-truth box of the \emph{target object}, i.e., the adversarial object placed in the scene, and any predicted box. The attack requires both a high \emph{victim miss rate}, the fraction of frames with $\mathrm{IoU}^*=0$ from the victim's ego view, and a high \emph{non-victim detection rate}, the fraction of frames with $\mathrm{IoU}^*>0$ from at least one non-victim ego view.

\emph{Defense-level ASR}: each defense exposes a different decision variable, so the success condition is defense-specific:
\begin{itemize}
    \item \textbf{CAD}: a frame succeeds when CAD reports a conflicted occupancy region that spatially matches the adversarial object's ground-truth region.
    \item \textbf{MATE}: a 10-frame trajectory succeeds when the victim's final agent trust falls below the track-count-specific threshold after the temporal trust update.
    \item \textbf{LUCIA}: a frame succeeds when, from a non-victim ego view, the victim's trust score is below $0.1$, i.e., it contributes less than 10\% to fusion.
    \item \textbf{MADE}: a frame succeeds when a non-victim ego marks the victim as the malicious collaborator.
    \item \textbf{ROBOSAC}: among paired observations where the benign case accepts a subgroup containing the victim, success means the attacked case accepts a subgroup that excludes the victim.
\end{itemize}

\emph{Average precision (AP)}: AP is computed after the corresponding fusion pipeline, at IoU $0.5$ on OPV2V and $0.3$ on V2X-Real, where the looser threshold keeps small box misalignments from V2X-Real's sensing noise and manual annotation from counting as complete misses. Fusion is unmodified when no defense is active. CAD, LUCIA, and MADE act per frame: CAD and MADE drop the marked collaborator from that frame's fusion, while LUCIA weights each collaborator by its current trust score. MATE instead updates agent trust over its 10-frame Bayesian window and uses the result as the fusion weight. No evaluated defense permanently removes a CAV.

\subsection{Attack Effectiveness}
\label{subsec:attack-effectiveness}

We evaluate perception discrepancy and defense impact on both benchmarks, then compare full OPV2V with the critical subsets to measure the benefit of deployment selection. Representative optimized objects appear in Appendix~\ref{app:optimized-meshes}. Across fusion protocols, backbones, and defenses, \attackname exposes a \textbf{general attribution weakness with scenario-dependent impact}: favorable geometry amplifies both victim penalization and utility loss.

\subsubsection{Perception-Level Discrepancy}
\label{subsubsec:perception-discrepancy}

\input{tables/tab_view_conditioned}

\emph{View-conditioned asymmetry persists across datasets and CP models.}
Across both datasets in Table~\ref{tab:view-conditioned-detection-results}, \textsc{FromReal} increases the victim miss rate while retaining substantial non-victim detectability, and \textsc{Hollow} follows the same but weaker trend. Late fusion attenuates this asymmetry but does not eliminate it. For intermediate fusion, removing the victim feature generally recovers non-victim detection on OPV2V and on AttFusion with V2X-Real, showing that the victim feature can propagate the local discrepancy. The less uniform Where2Comm results on V2X-Real indicate that this effect depends on the fusion architecture and sensing conditions. Despite different benign baselines, \textsc{FromReal} and \textsc{Hollow} remain clearly separated from benign victim-miss rates while retaining substantial non-victim detection on both datasets.

\emph{PointPillars is more tolerant of the small \textsc{Attached} occluder.}
\textsc{Attached} strongly separates the two views on PIXOR but is weaker on PointPillars and V2X-Real. A plain cuboid, an unoptimized box of car-like size used as a shape-prior-free reference, confirms this detector-specific behavior: its victim miss and non-victim detection rates are $63.56\%/41.07\%$ on PointPillars and $81.88\%/21.96\%$ on PIXOR. Because \textsc{Attached} leaves the host vehicle largely intact, PointPillars more readily preserves the vehicle hypothesis from the victim view.

\emph{Shape initialization is critical for view-conditioned asymmetry.}
A cuboid optimized jointly for both view-conditioned objectives, with no shape prior to resolve either side in advance, achieves a $73.19\%$ victim miss rate but only a $37.10\%$ non-victim detection rate. This imbalance is consistent with opposing updates on shared vertices: suppressing victim-view evidence also damages the geometry needed for non-victim detection. It motivates \textsc{FromReal}'s localized vertex optimization, which limits changes to the victim-facing region.

\subsubsection{Defense-Level Trust Poisoning}
\label{subsubsec:defense-level-poisoning}

\input{tables/tab_defense_results}

\emph{\attackname most effectively poisons defenses that react to localized inconsistency.}
The rows without \defensename in Table~\ref{tab:defense-results} show that \textsc{FromReal} is most effective against MATE and MADE, while \textsc{Hollow} is most effective against CAD and in LUCIA's strongest case. MATE, CAD, and MADE react strongly enough to the localized inconsistency to penalize the victim. LUCIA remains closer to its benign baseline except for \textsc{Hollow} on Where2Comm, indicating limited sensitivity to point-cloud-stage single-object errors rather than a strong trust-poisoning surface. Section~\ref{subsubsec:lucia-effectiveness} examines this behavior. MATE exhibits a smaller attack--benign ASR gap on V2X-Real than on OPV2V. Section~\ref{subsubsec:mate-effectiveness} explains how real-world tracking noise interacts with its temporal evidence accumulation.

\emph{Defense ASR does not necessarily predict AP loss.}
In an idealized ablation that deletes only the victim's detection of the target object before MATE runs, ASR reaches $98.31\%$ while AP stays at $84.45\%$, showing that crossing MATE's trust threshold does not imply proportional utility loss. The matching ablation for MADE, which instead deletes the target's points from the victim's cloud before feature extraction, yields $67.43\%$ ASR and $80.44\%$ AP, consistent with binary collaborator exclusion coupling ASR more directly to AP. The corresponding LUCIA result is discussed in Figure~\ref{subfig:lucia-ideal-discrepancy}.

\begin{table}[t]
    \centering
    \caption{Reference AP baselines on OPV2V. \textsc{Benign} uses normal fusion, while \textsc{Victim Excluded} removes the victim from collaboration in all frames. Values are in \%.}
    \label{tab:ap-baselines}
    \input{tables/ap_baselines}
\end{table}

\emph{Excluding the victim in every frame bounds the AP loss any attack can cause.}
Table~\ref{tab:ap-baselines} compares benign fusion with a reference that excludes the victim in every frame. This maximal suppression upper-bounds AP loss from trust poisoning and, because both conditions use benign scenes, isolates the cost of losing the victim from effects of the adversarial object. The bound depends on the victim's unique coverage, ranging from $4.83$ points on PIXOR-Late to $10.61$ points on PointPillars-Attentive.

\subsubsection{Impact of Attack Parameters}
\label{subsubsec:attack-parameters}

We analyze how the vertex mask, during adversarial object optimization, mediates the gradient conflict between victim and non-victim losses for \fromreal, as mentioned in Section~\ref{sec:design}.

\emph{View separation reduces gradient conflict.}
For \textsc{FromReal}, the vertex mask $\mathcal{M}$ determines how much geometry can be changed to hide the object from the victim, but also how much benign geometry may be perturbed from non-victim views. To quantify this tradeoff, we count the vertices in $\mathcal{M}$ whose victim-loss and non-victim-loss gradients oppose each other at a non-victim angle $\theta$:
\[
N_{\mathrm{conf}}(\theta)=
\left|
\left\{
i\in\mathcal{M}:
\cos(g^i_v,g^i_n(\theta))<0
\right\}
\right|,
\]
where $g^i_v$ and $g^i_n(\theta)$ are the gradients on vertex $i$ induced by the victim and non-victim losses.

\begin{figure}[t]
    \centering
    \includegraphics[width=0.8\columnwidth]{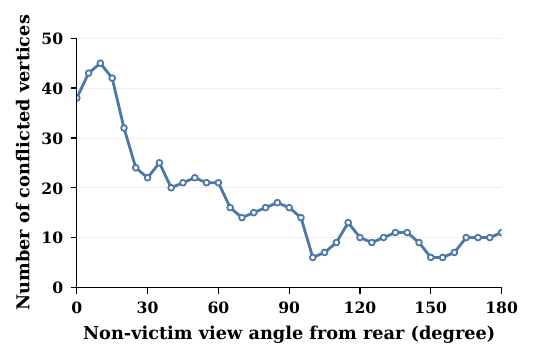}
    \caption{Gradient-conflict trend for \textsc{FromReal}. The y-axis counts mask vertices whose victim and non-victim loss gradients point in opposing directions.}
    \label{fig:vertex-conflict-tradeoff}
\end{figure}

Figure~\ref{fig:vertex-conflict-tradeoff} shows that conflict is highest when the non-victim observes the object from a view close to the victim's view. Around $\ang{20}$, the count drops sharply as the non-victim leaves the vulnerable cone and shares fewer victim-visible vertices. Beyond side-view angles, the curve stays low and fluctuates around a small nonzero value because the remaining overlap is mainly on the rooftop surface. This supports the \textsc{FromReal} design choice in Section~\ref{subsec:offline-optimization}: optimizing a set of localized vertices gives it enough perturbable geometry for disappearance from the victim while limiting conflict with normal detection from other non-victim vehicles.

\input{tables/deployment_robustness}

\emph{\attackname remains effective under imperfect object alignment, supporting practical deployment.}
Table~\ref{tab:deployment-robustness} shows that \attackname preserves view asymmetry under random heading errors of up to $\ang{10}$: victim-view suppression remains effective, non-victim detection stays high, and the discrepancy continues to activate defenses after fusion. \textsc{Hollow} is more sensitive because yaw shifts victim rays from its open channel onto its side panels, revealing more of the structure. Overall, \attackname tolerates moderate placement error without per-scene mesh retuning.

\subsubsection{Impact of Scenario Selection}
\label{subsubsec:scenario-impact}

\begin{figure*}[!t]
    \centering
    \includegraphics[width=\textwidth]{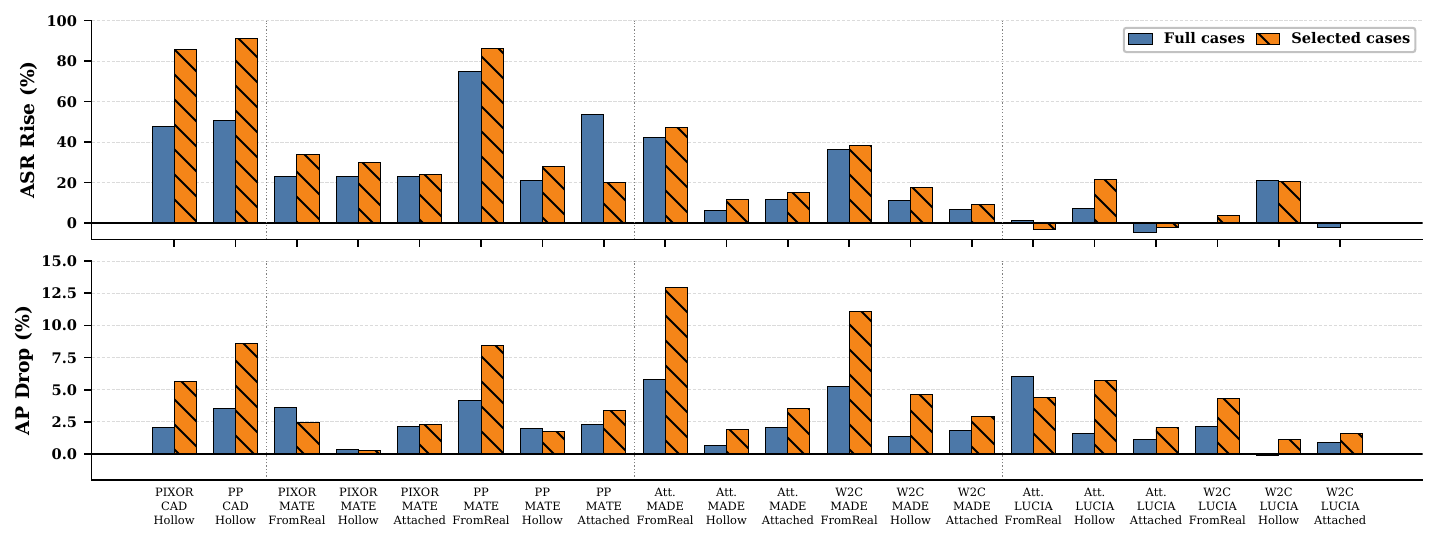}
    \caption{Effect of scenario selection on outcomes of attack on CP defenses (w/o \defensename). Bars compare full-case deltas with selected critical-case deltas, measured as attack--benign ASR (top) and benign--attack AP (bottom). PP, Att., and W2C denote PointPillars, AttFusion, and Where2Comm, respectively.}
    \label{fig:selected-case-delta-comparison}
\end{figure*}

\emph{Scenario selection amplifies both defense-level ASR and AP loss.}
The full-dataset protocol uses a random victim and constrains non-victim geometry mainly by visibility. Selecting cases with larger cross-view discrepancy and stronger victim contribution increases both defense-level ASR and AP loss for most model--defense combinations in Figure~\ref{fig:selected-case-delta-comparison}. This validates Insight~\ref{ins:scenario}: deployment geometry amplifies both victim penalization and downstream utility loss.

\subsection{Analysis of CP Defenses}
\label{subsec:impact-factor}

Analyzing how existing CP defenses respond to \attackname, we identify their \textbf{sensitivity-attribution dilemma}.
Across the defenses we study, robustness to \attackname tends to track insensitivity to localized single-object errors. Defenses that respond strongly to such an error are generally easier to turn against a benign victim, while those that smooth it away resist the attack but retain less signal for detecting genuine single-object fabrication. A low ASR therefore does not by itself indicate that a defense attributes the inconsistency correctly.

Specifically, we examine this trade-off through MATE's track count, LUCIA's compression ratio, and ROBOSAC's consensus threshold. We then use additional LUCIA diagnostics to explain why its response to point-cloud-stage discrepancies is limited.

\subsubsection{On Late-Fusion Consistency Check (MATE)}
\label{subsubsec:mate-effectiveness}

We study how MATE's number of tracks shapes its susceptibility to \attackname. MATE~\cite{hallyburton2025security} is a late-fusion defense that estimates agent trust through Bayesian updates over trust pseudomeasurements (PSMs): matched track pairs between ego agent and aggregator supply positive evidence, while aggregator tracks that should be visible to an agent but are missing from its report supply negative evidence. Because MATE accumulates PSMs over ten frames, trust poisoning depends on a stable target track. Noisy detections and track associations in V2X-Real can interrupt target-specific negative evidence while introducing unrelated benign mismatches. This raises MATE's benign false-positive rate and reduces the attack--benign separation.

\emph{Track count balances benign and negative evidence.}
We report MATE ASR over observed $1$, $3$, $5$, and $10$ tracks. Because MATE's trust scale grows with the number of matched tracks, the corresponding agent-trust thresholds are set to $0.15$, $0.35$, $0.40$, and $0.45$. More tracks provide additional positive PSMs that absorb negative evidence, but they also produce more negative PSMs when the victim disagrees with non-victims.

\begin{figure}[t]
    \centering
    \begin{minipage}[t]{0.49\columnwidth}
        \centering
        \includegraphics[width=\linewidth]{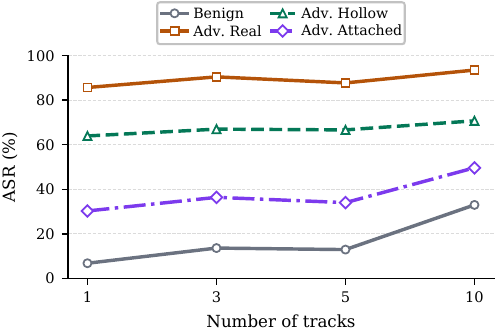}
        \caption{Impact of the number of tracks on MATE.}
        \label{fig:mate-track-count}
    \end{minipage}
    \hfill
    \begin{minipage}[t]{0.49\columnwidth}
        \centering
        \includegraphics[width=\linewidth]{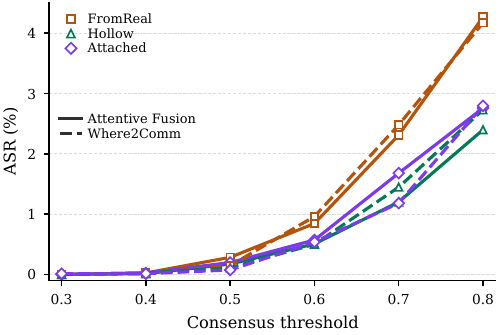}
        \caption{Impact of consensus thresholds on ROBOSAC.}
        \label{fig:robosac-threshold-sensitivity}
    \end{minipage}
\end{figure}

Figure~\ref{fig:mate-track-count} reveals two competing effects. More tracks supply more positive PSMs from benign matches, raising agent trust and absorbing isolated errors. They also create more visibility-versus-report checks: when \attackname repeatedly makes the victim disagree with non-victims on the adversarial object, the victim accumulates negative PSMs and fails to rebuild trust at the same rate as benign agents. The mesh-dependent gap reflects the strength of the physical inconsistency: \textsc{FromReal} produces the strongest MATE ASR curve, \textsc{Hollow} stays above the benign baseline, and \textsc{Attached} is weaker but visible. \attackname therefore exploits MATE by turning one localized physical discrepancy into repeated negative trust evidence.

\subsubsection{On Intermediate-Feature Consistency (LUCIA)}
\label{subsubsec:lucia-effectiveness}

LUCIA~\cite{wang2025threat} converts pairwise L1 distances between pooled, normalized BEV features into collaborator trust. We examine its compression ratio (CR) and feature-distance metric.

\begin{figure}[!t]
    \centering
    \begin{subfigure}[t]{0.48\columnwidth}
        \centering
        \includegraphics[width=\linewidth]{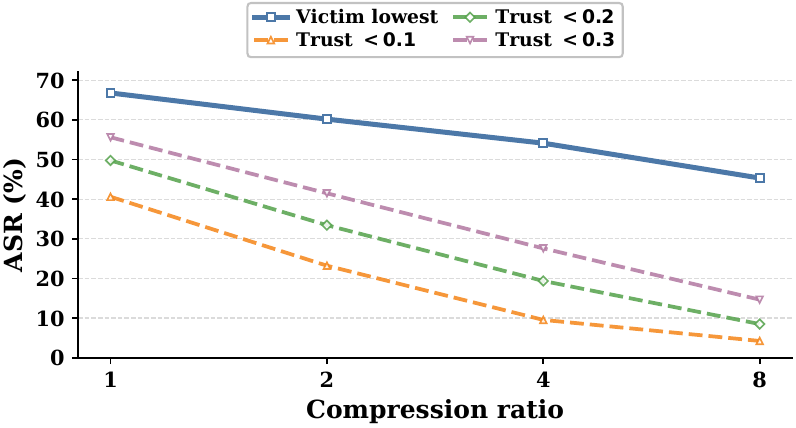}
        \caption{Compression ratio}
        \label{subfig:lucia-compression-ratio}
    \end{subfigure}
    \hfill
    \begin{subfigure}[t]{0.48\columnwidth}
        \centering
        \includegraphics[width=\linewidth]{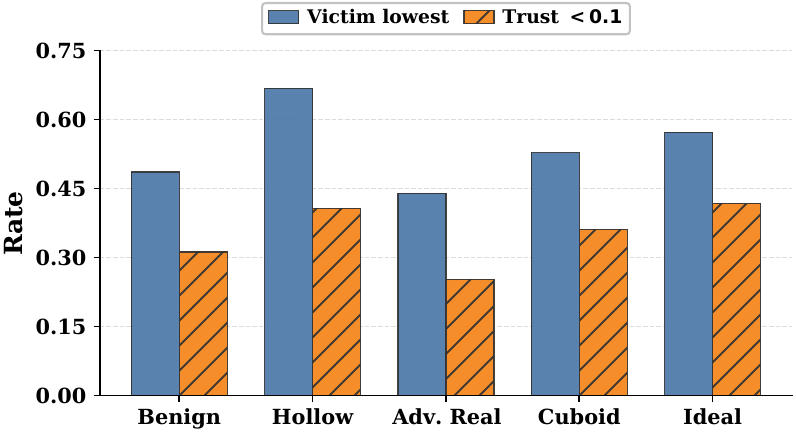}
        \caption{Idealized visibility discrepancy}
        \label{subfig:lucia-ideal-discrepancy}
    \end{subfigure}
    \begin{subfigure}[t]{0.48\columnwidth}
        \centering
        \includegraphics[width=\linewidth]{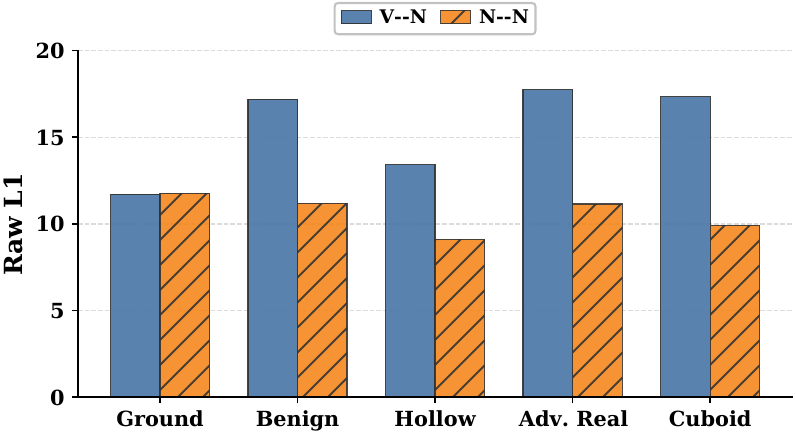}
        \caption{Same object across different views}
        \label{subfig:lucia-cross-view-inconsistency}
    \end{subfigure}
    \hfill
    \begin{subfigure}[t]{0.48\columnwidth}
        \centering
        \includegraphics[width=\linewidth]{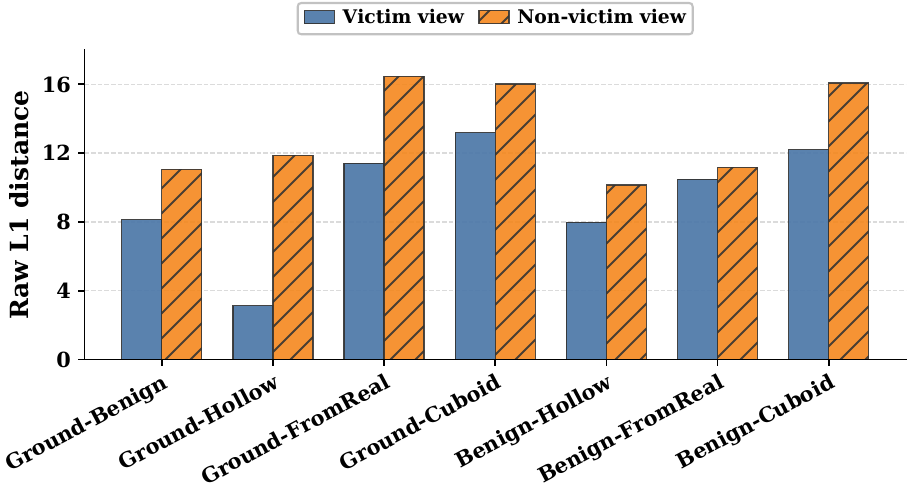}
        \caption{Different objects under the same view}
        \label{subfig:lucia-same-view-object-inconsistency}
    \end{subfigure}
    \caption{Impacting factors and diagnostics for LUCIA.}
    \label{fig:lucia-analysis}
\end{figure}

\emph{A larger compression ratio buys robustness only by discarding the evidence LUCIA relies on.}
Figure~\ref{subfig:lucia-compression-ratio} shows that ASR falls as CR grows. Larger pooling windows average the object-local discrepancy with unaffected context. This resists \attackname by smoothing away the same fine-grained inconsistency evidence that LUCIA uses for detection.

\emph{Where2Comm's spatial selection concentrates the discrepancy, raising LUCIA ASR above AttFusion.}
Where2Comm's confidence-guided mask suppresses low-information background cells, so target cells occupy more of LUCIA's compared region and their mismatch is less diluted. AttFusion's dense BEV attention retains more unaffected context before feature distance is computed.

\emph{Even perfect victim-view object removal barely raises LUCIA's suspicion.}
We replace the victim's target points with ground returns while leaving the object intact for non-victims, the strongest disappearance case in our threat model. In Figure~\ref{subfig:lucia-ideal-discrepancy}, ASR remains only moderately above benign and comparable to \textsc{Hollow} and cuboid, while \textsc{FromReal} is no more suspicious than the benign car. Feature encoding, pooling, and normalization therefore attenuate even perfect victim-view removal.

\emph{LUCIA's feature distance does not track physical visibility.}
We measure normalized local-feature L1 distance for the same object across views and for different objects under one view. Figures~\ref{subfig:lucia-cross-view-inconsistency} and~\ref{subfig:lucia-same-view-object-inconsistency} show that \textsc{Hollow}'s ground-like victim view does not produce the largest cross-view distance, and ordinary viewpoint changes can be comparable. Unlike broad digital feature perturbations~\cite{wang2025threat}, \attackname changes sparse returns before feature extraction, so encoding and pooling do not amplify its ray-level asymmetry into a unique feature outlier.

\subsubsection{On Consensus-Based Defense (ROBOSAC)}
\label{subsubsec:robosac-effectiveness}

ROBOSAC~\cite{li2023among} accepts a sampled collaborator subgroup when the Jaccard similarity between its fused bounding-box set and the ego-only prediction set exceeds a consensus threshold.

\emph{A single-object error rarely changes ROBOSAC's scene-level consensus.}
Figure~\ref{fig:robosac-threshold-sensitivity} shows that ASR remains low across the evaluated consensus thresholds, backbones, and shape priors, although stricter consensus modestly increases sensitivity to the attack. Because \attackname changes only one object, unaffected detections preserve the similarity of the complete bounding-box sets, and subgroups containing the victim generally remain accepted. ROBOSAC therefore marks a scope boundary for \attackname: its scene-level decision usually absorbs the localized discrepancy instead of penalizing the victim.

\subsection{Mitigation Effectiveness}
\label{subsec:defense-effectiveness}

We evaluate how \defensename reduces \attackname's defense-level ASR and restores AP.

\emph{\defensename restores AP when defenses would otherwise exclude the misidentified victim.}
Table~\ref{tab:defense-results} compares the defense-aware results without \defensename, discussed in Section~\ref{subsec:attack-effectiveness}, against those obtained with \defensename for each evaluated defense.

The ASR and AP differences show that self-reflection reduces incorrect victim penalization across all evaluated defenses. For CAD and MADE, the effect is direct. Both exclude the collaborator they mark, so preventing the victim from being marked restores its useful contribution and improves AP. For LUCIA, \defensename reduces the ASR to a near-benign self-reflection level and improves AP, showing that masking locally inconsistent regions can stabilize feature-consistency weighting even when the original LUCIA signal is not strongly effective.

MATE remains the main exception in AP behavior for the same reason discussed in Section~\ref{subsec:attack-effectiveness}: trust is a continuous fusion weight, so AP is not determined solely by whether the victim crosses the ASR threshold.

\subsection{Physical Experiment}
\label{subsubsec:physical-experiment}

\emph{The shape priors induce view-conditioned detection despite imperfect fabrication.}
High-fidelity, full-size fabrication of the optimized meshes requires industrial manufacturing. We instead manually build full-size prototypes of the initial \textsc{Hollow} and \textsc{Attached} geometries to test whether their physical principles survive practical construction errors before gradient refinement. The \textsc{Hollow} prototype uses two rigid, sedan-scale side panels stabilized by thin supports whose cross-sections are below the LiDAR's angular resolution. The \textsc{Attached} prototype mounts a planar occluder behind a host vehicle using an adjustable strut. We process their rear- and side-view captures with a pretrained PointPillars late-fusion model.

\begin{figure}[!t]
    \centering
    \begin{subfigure}[t]{\columnwidth}
        \centering
        \includegraphics[width=\linewidth]{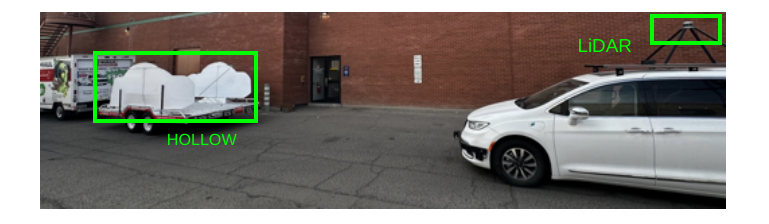}
        \caption{Full-size \textsc{Hollow}.}
        \label{subfig:hollow-location}
    \end{subfigure}
    \\[2pt]
    \begin{subfigure}[t]{\columnwidth}
        \centering
        \includegraphics[width=\linewidth]{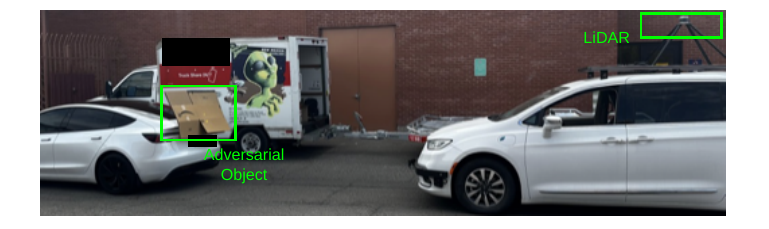}
        \caption{Full-size \textsc{Attached}.}
        \label{subfig:attached-location}
    \end{subfigure}
    \\[2pt]
    \begin{subfigure}[t]{\columnwidth}
        \centering
        \includegraphics[width=0.9\linewidth]{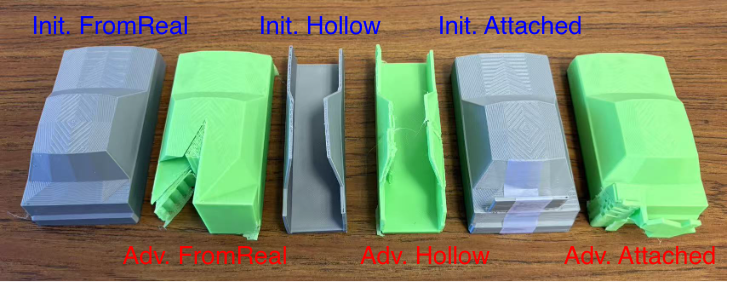}
        \caption{Initial and optimized objects 3D-printed at 1:35 scale.}
        \label{subfig:miniature-prototypes}
    \end{subfigure}
    \caption{Physical fabrication of full-size and miniature \attackname objects.}
    \label{fig:physical-prototypes}
\end{figure}

\emph{The optimized adversarial geometries are physically coherent and 3D-printable.}
We additionally fabricate the initial and optimized versions of all three priors as 1:35-scale 3D prints. Their successful fabrication shows that the constrained optimization preserves physically producible geometry rather than generating purely numerical, non-manufacturable shapes. Figure~\ref{fig:physical-prototypes} documents both fabrication studies.

\begin{figure}[!t]
    \centering
    \begin{subfigure}[t]{0.48\columnwidth}
        \centering
        \includegraphics[width=\linewidth]{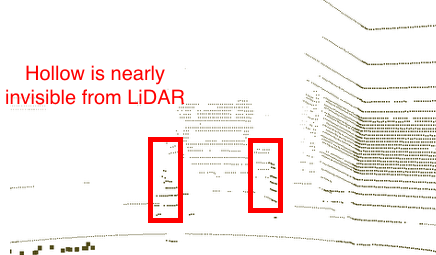}
        \caption{\textsc{Hollow}, rear}
        \label{subfig:physical-hollow-rear}
    \end{subfigure}
    \hfill
    \begin{subfigure}[t]{0.48\columnwidth}
        \centering
        \includegraphics[width=\linewidth]{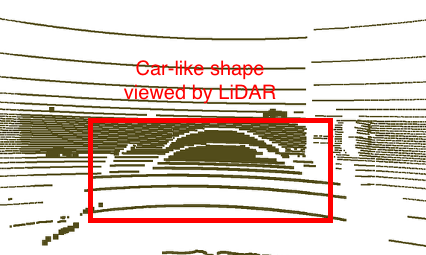}
        \caption{\textsc{Hollow}, side}
        \label{subfig:physical-hollow-side}
    \end{subfigure}
    \\
    \vspace{2pt}
    \begin{subfigure}[t]{0.48\columnwidth}
        \centering
        \includegraphics[width=\linewidth]{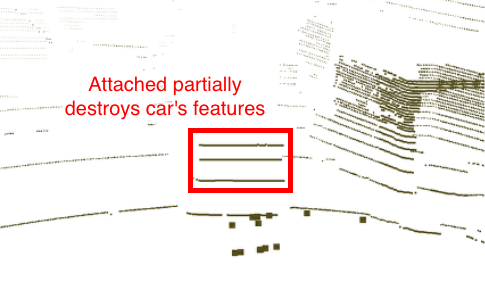}
        \caption{\textsc{Attached}, rear}
        \label{subfig:physical-attached-rear}
    \end{subfigure}
    \hfill
    \begin{subfigure}[t]{0.48\columnwidth}
        \centering
        \includegraphics[width=\linewidth]{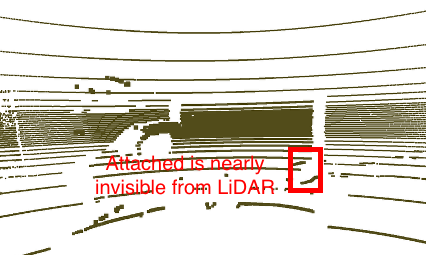}
        \caption{\textsc{Attached}, side}
        \label{subfig:physical-attached-side}
    \end{subfigure}
    \\
    \vspace{2pt}
    \begin{subfigure}[t]{0.32\columnwidth}
        \centering
        \includegraphics[width=\linewidth]{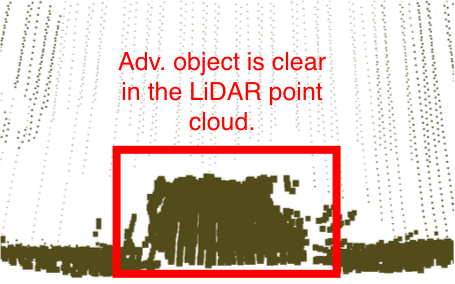}
        \caption{Adv. \textsc{FromReal}}
        \label{subfig:miniature-fromreal-pcd}
    \end{subfigure}
    \hfill
    \begin{subfigure}[t]{0.32\columnwidth}
        \centering
        \includegraphics[width=\linewidth]{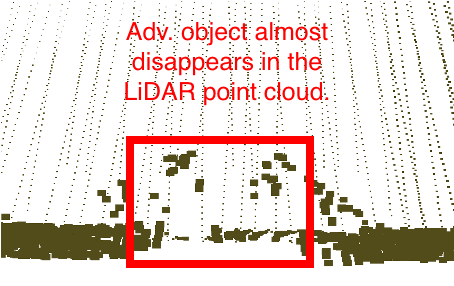}
        \caption{Adv. \textsc{Hollow}}
        \label{subfig:miniature-hollow-pcd}
    \end{subfigure}
    \hfill
    \begin{subfigure}[t]{0.32\columnwidth}
        \centering
        \includegraphics[width=\linewidth]{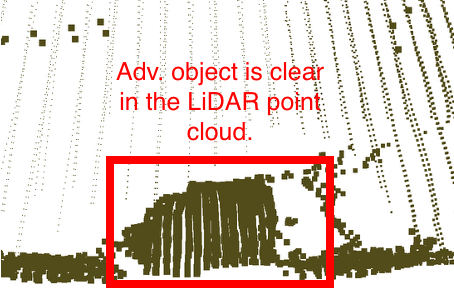}
        \caption{Adv. \textsc{Attached}}
        \label{subfig:miniature-attached-pcd}
    \end{subfigure}
    \caption{Real LiDAR returns from full-size prototypes (a--d) and optimized 1:35-scale miniatures (e--g). The full-size captures compare vulnerable rear views ($\Theta_v$) with non-vulnerable side views ($\Theta_n$). All miniature captures use the rear view.}
    \label{fig:physical-point-clouds}
\end{figure}

Figure~\ref{fig:physical-point-clouds} shows consistent sensor-level behavior across both scales. For the full-size \textsc{Hollow}, the rear view produces almost no panel returns, so PointPillars does not detect a vehicle, while the side view forms a dense vehicle-like profile that it detects. For full-size \textsc{Attached}, the rear occluder suppresses the host's characteristic geometry and causes PointPillars to miss it, while the thin side profile preserves detection. Thus, the intended asymmetry survives imperfect object crafting. From the rear of the optimized miniatures, \textsc{FromReal} and \textsc{Attached} produce clear object profiles, whereas \textsc{Hollow} returns almost no points because rays pass through its open channel. Together, the full-size detections validate robustness to construction error, while the miniature prints and scans demonstrate that the optimized shapes are physically producible and preserve their intended ray-level behavior.

%% file: tables/tab_view_conditioned.tex
\begin{table*}[!t]\centering\small
\caption{View-conditioned target detection on OPV2V and V2X-Real. Numbers denote rate of missed detection on the adversarial object for the victim agent (higher is better), and the rate of correct detection on the adversarial object for non-victim agents (higher is better), respectively. \emph{Single}/\emph{Fused} denote ego-view/late-fused outputs. \emph{w/ Victim} and \emph{w/o Victim} denotes including and excluding the victim feature in the fusion, respectively. }\label{tab:view-conditioned-detection-results}
\resizebox{\textwidth}{!}{
\begin{tabular}{ll||rr|rr|rr|rr||rr|rr|rrr}\toprule
\multicolumn{2}{c||}{\textbf{Dataset}} &\multicolumn{8}{c||}{\textbf{OPV2V}} &\multicolumn{6}{c}{\textbf{V2X-Real}} \\\cmidrule{1-16}
\multicolumn{2}{c||}{\textbf{Fusion Method}} &\multicolumn{4}{c}{\textbf{Late Fusion}} &\multicolumn{4}{|c||}{\textbf{Intermediate Fusion}} &\multicolumn{2}{c}{\textbf{Late Fusion}} &\multicolumn{4}{|c}{\textbf{Intermediate Fusion}} \\\cmidrule{1-16}
\multicolumn{2}{c||}{\textbf{Backbone}} &\multicolumn{2}{c}{\textbf{PIXOR}} &\multicolumn{2}{|c}{\textbf{PointPillars}} &\multicolumn{2}{|c}{\textbf{AttFusion}} &\multicolumn{2}{|c||}{\textbf{Where2Comm}} &\multicolumn{2}{c}{\textbf{PointPillars}} &\multicolumn{2}{|c}{\textbf{AttFusion}} &\multicolumn{2}{|c}{\textbf{Where2Comm}}\\\cmidrule{1-16}
\textbf{View} &\textbf{Mesh} &\emph{Single} &\emph{Fused} &\emph{Single} &\emph{Fused} &\emph{w/ Victim} &\emph{w/o Victim} &\emph{w/ Victim} &\emph{w/o Victim} &\emph{Single} &\emph{Fused} &\emph{w/ Victim} &\emph{w/o Victim} &\emph{w/ Victim} &\emph{w/o Victim} \\\midrule
\rowcolor{LightGreen}
\cellcolor{white} & Benign &20.03 &4.00 &9.16 &3.19 &7.53 &-- &19.78 &-- &33.26 &13.29 &40.36 &-- &29.85 &-- \\
\rowcolor{Gray}\cellcolor{white} &Adv. \textsc{FromReal} &88.50 &30.33 &\textbf{90.81} &\textbf{24.44} &\textbf{34.72} &-- &\textbf{63.56} &-- &\textbf{78.02} &\textbf{25.80} &\textbf{69.09} &-- &\textbf{58.85} &-- \\
&Adv. \textsc{Attached} &\textbf{89.59} &\textbf{31.56} &27.67 &9.47 &12.82 &-- &34.62 &-- &35.27 &14.39 &44.55 &-- &29.54 &-- \\
\rowcolor{Gray}\cellcolor{white}\multirow{-4}{*}{Victim} &Adv. \textsc{Hollow} &84.68 &22.74 &60.56 &13.09 &23.73 &-- &41.52 &-- &56.67 &18.76 &54.86 &-- &52.21 &-- \\
\midrule
\rowcolor{LightGreen}
\cellcolor{white} & Benign &71.88 &86.46 &74.55 &87.15 &88.00 &86.16 &79.84 &81.40 &72.47 &77.04 &62.60 &66.85 &75.15 &72.96 \\
\rowcolor{Gray}\cellcolor{white} &Adv. \textsc{FromReal} &51.47 &66.98 &61.38 &73.12 &68.88 &75.67 &51.44 &59.98 &62.92 &69.65 &50.10 &60.29 &62.32 &63.31 \\
&Adv. \textsc{Attached} &51.57 &67.02 &\textbf{67.18} &\textbf{83.19} &\textbf{82.48} &\textbf{81.98} &66.17 &71.54 &\textbf{72.07} &\textbf{77.00} &\textbf{61.83} &\textbf{65.89} &\textbf{73.64} &\textbf{70.80} \\
\rowcolor{Gray}\cellcolor{white}\multirow{-4}{*}{Non-Victim} &Adv. \textsc{Hollow} &\textbf{54.89} &\textbf{73.01} &63.98 &80.61 &74.73 &79.55 &\textbf{67.30} &\textbf{73.59} &62.99 &72.74 &58.13 &62.37 &63.07 &62.23 \\
\bottomrule
\end{tabular}
}
\end{table*}

%% file: tables/tab_defense_results.tex
\begin{table*}[!t]\centering\scriptsize
\caption{Defense-level attack and mitigation results on OPV2V and V2X-Real. \texttt{TR} denotes \defensename. N gives results without TR, and Y gives $\Delta=\text{with TR}-\text{without TR}$. Higher ASR and lower AP indicate stronger attacks. Lower $\Delta$ASR and higher $\Delta$AP indicate stronger mitigation. Bold and underline mark the strongest attack and mitigation among adversarial meshes. AP uses IoU 0.5/0.3 on OPV2V/V2X-Real. Values are in \%.}
    \label{tab:defense-results}
\resizebox{\textwidth}{!}{
\begin{tabular}{cr||rr|rr|rr|rr||rr|rr|rr|rr||rr|rr|rr|rr}\toprule
\multicolumn{2}{c||}{\textbf{Dataset}} &\multicolumn{16}{c||}{\textbf{OPV2V}} &\multicolumn{8}{c}{\textbf{V2X-Real}} \\\cmidrule{1-26}
\multicolumn{2}{c||}{\textbf{Fusion Method}} &\multicolumn{8}{c}{\textbf{Late Fusion}} &\multicolumn{8}{|c||}{\textbf{Intermediate Fusion}} &\multicolumn{2}{c}{\textbf{Late Fusion}} &\multicolumn{6}{|c}{\textbf{Intermediate Fusion}} \\\cmidrule{1-26}
\multicolumn{2}{c||}{\textbf{Backbone}} &\multicolumn{4}{c}{\textbf{PIXOR}} &\multicolumn{4}{|c}{\textbf{PointPillars}} &\multicolumn{4}{|c}{\textbf{AttFusion}} &\multicolumn{4}{|c||}{\textbf{Where2Comm}} &\multicolumn{2}{c}{\textbf{PointPillars}} &\multicolumn{4}{|c}{\textbf{AttFusion}} &\multicolumn{2}{|c}{\textbf{Where2Comm}} \\\cmidrule{1-26}
\multicolumn{2}{c||}{\textbf{Defense}} &\multicolumn{2}{c}{\textbf{MATE}} &\multicolumn{2}{|c}{\textbf{CAD}} &\multicolumn{2}{|c}{\textbf{MATE}} &\multicolumn{2}{|c}{\textbf{CAD}} &\multicolumn{2}{|c}{\textbf{LUCIA}} &\multicolumn{2}{|c}{\textbf{MADE}} &\multicolumn{2}{|c}{\textbf{LUCIA}} &\multicolumn{2}{|c||}{\textbf{MADE}} &\multicolumn{2}{c}{\textbf{MATE}} &\multicolumn{2}{|c}{\textbf{LUCIA}} &\multicolumn{2}{|c}{\textbf{MADE}} &\multicolumn{2}{|c}{\textbf{LUCIA}} \\\cmidrule{1-26}
\textbf{Mesh} &\texttt{TR}? &\textbf{ASR} &\textbf{AP} &\textbf{ASR} &\textbf{AP} &\textbf{ASR} &\textbf{AP} &\textbf{ASR} &\textbf{AP} &\textbf{ASR} &\textbf{AP} &\textbf{ASR} &\textbf{AP} &\textbf{ASR} &\textbf{AP} &\textbf{ASR} &\textbf{AP} &\textbf{ASR} &\textbf{AP} &\textbf{ASR} &\textbf{AP} &\textbf{ASR} &\textbf{AP} &\textbf{ASR} &\textbf{AP} \\\midrule
\rowcolor{LightGreen} &N &19.73 &76.98 &-- &-- &12.93 &87.14 &-- &-- &31.74 &80.84 &9.89 &87.43 &36.56 &78.64 &9.86 &83.37 &27.78 &90.07 &29.62 &72.65 &9.28 &78.74 &30.17 &66.30 \\

\rowcolor{LightGreen}\multirow{-2}{*}{Benign} &Y &-5.44 &+0.55 &-- &-- &-3.41 &-0.14 &-- &-- &-22.32 &+2.11 &-3.08 &+0.08 &-25.63 &+1.74 &-3.52 &+0.11 &-11.11 &-0.28 &-20.40 &+2.62 &-2.43 &+0.87 &-21.80 &+2.79 \\
\midrule
\rowcolor{Gray}
\multirow{2}{*}{\cellcolor{white}\makecell{\cellcolor{white}Adv.\\\cellcolor{white}\textsc{FromReal}}} &N &\textbf{42.86} &\textbf{73.33} &-- &-- &\textbf{87.76} &\textbf{82.96} &-- &-- &32.90 &\textbf{74.82} &\textbf{52.14} &\textbf{81.61} &37.03 &\textbf{76.49} &\textbf{45.99} &\textbf{78.11} &\textbf{55.16} &86.96 &31.41 &\textbf{71.78} &\textbf{44.38} &\textbf{76.40} &30.91 &\textbf{65.38} \\
&Y &-20.41 &+0.24 &-- &-- &\underline{-59.19} &\underline{+0.94} &-- &-- &-23.26 &\underline{+5.86} &\underline{-42.97} &\underline{+1.65} &-25.40 &+1.66 &\underline{-34.95} &\underline{+1.31} &\underline{-30.56} &+0.31 &-22.89 &+2.74 &\underline{-34.35} &\underline{+2.51} &-22.57 &\underline{+2.84} \\
\midrule
\rowcolor{Gray}
\multirow{2}{*}{\cellcolor{white}\makecell{\cellcolor{white}Adv.\\\cellcolor{white}\textsc{Attached}}} &N &\textbf{42.86} &74.85 &-- &-- &34.01 &84.81 &-- &-- &27.11 &79.70 &21.60 &85.34 &34.67 &77.71 &16.78 &81.51 &34.13 &88.82 &30.39 &72.41 &19.46 &78.38 &30.00 &66.17 \\
&Y &-18.37 &-0.20 &-- &-- &-18.02 &-0.09 &-- &-- &-19.17 &+2.04 &-10.08 &+0.33 &-24.84 &\underline{+1.77} &-10.77 &+0.37 &-15.48 &-0.60 &-19.61 &+2.61 &-7.89 &+1.29 &-22.16 &+2.72 \\
\midrule
\rowcolor{Gray}
\multirow{2}{*}{\cellcolor{white}\makecell{\cellcolor{white}Adv.\\\cellcolor{white}\textsc{Hollow}}} &N &\textbf{42.86} &76.65 &47.65 &64.45 &66.67 &85.12 &50.84 &82.29 &\textbf{39.09} &79.22 &16.33 &86.72 &\textbf{57.48} &78.71 &20.88 &81.99 &44.44 &\textbf{86.00} &\textbf{40.33} &72.18 &24.53 &77.97 &\textbf{39.79} &65.82 \\
&Y &\underline{-21.09} &\underline{+0.25} &-47.18 &+2.14 &-44.22 &-1.11 &-50.33 &+3.17 &\underline{-30.79} &+2.56 &-5.87 &+0.19 &\underline{-47.21} &+0.83 &-11.98 &+0.33 &-26.19 &\underline{+0.60} &\underline{-31.99} &\underline{+2.93} &-13.59 &+1.88 &\underline{-29.51} &+2.73 \\
\bottomrule
\end{tabular}}
\end{table*}

%% file: tables/ap_baselines.tex
\scriptsize
\setlength{\tabcolsep}{4pt}
\renewcommand*{\arraystretch}{1.2}
\resizebox{0.8\linewidth}{!}{
    \begin{tabular}{@{}lccc@{}}
        \toprule
        Model & \textsc{Benign} & \textsc{Victim Excluded} & Drop \\
        \midrule
        PIXOR-Late & 66.56 & 61.73 & 4.83 \\
        \hline
        PointPillars-Late & 85.83 & 75.44 & 10.39 \\
        \hline
        PointPillars-Attentive & 88.60 & 77.99 & 10.61 \\
        \hline
        PointPillars-Where2Comm & 84.74 & 75.21 & 9.53 \\
        \bottomrule
    \end{tabular}
}

%% file: tables/deployment_robustness.tex
\begin{table}[t]
    \centering
    \caption{Deployment robustness on OPV2V under random object-yaw bias. Defense ASR is measured without \defensename. Values are in \%.}
    \label{tab:deployment-robustness}
    \scriptsize
    \setlength{\tabcolsep}{1.8pt}

    \begin{subtable}[t]{\columnwidth}
        \centering
        \caption{Target detection.}
        \label{tab:deployment-robustness-detection}
        \resizebox{0.7\linewidth}{!}{%
        \begin{tabular}{@{}llcccc@{}}
            \toprule
            & & \multicolumn{2}{c}{Late Fusion} & \multicolumn{2}{c}{Intermediate Fusion} \\
            \cmidrule(lr){3-4}\cmidrule(lr){5-6}
            Adv. Mesh & View & \emph{Single} & \emph{Fused} & \emph{w/ Victim} & \emph{w/o Victim} \\
            \midrule
            \multirow{2}{*}{\textsc{FromReal}} & Victim & 77.74 & 22.30 & 29.98 & -- \\
                                                     & Non-Victim & 60.98 & 80.68 & 70.57 & 75.05 \\
            \midrule
            \multirow{2}{*}{\textsc{Attached}} & Victim & 23.00 & 8.30 & 12.14 & -- \\
                                                     & Non-Victim & 67.11 & 90.87 & 83.02 & 81.82 \\
            \midrule
            \multirow{2}{*}{\textsc{Hollow}} & Victim & 40.21 & 10.30 & 17.19 & -- \\
                                                   & Non-Victim & 63.09 & 89.01 & 78.59 & 78.28 \\
            \bottomrule
        \end{tabular}
        }
    \end{subtable}
    \par\smallskip
    \begin{subtable}[t]{\columnwidth}
        \centering
        \caption{Defense ASR.}
        \label{tab:deployment-robustness-defenses}
        \resizebox{0.52\linewidth}{!}{%
        \begin{tabular}{@{}lcccc@{}}
            \toprule
            Adv. Mesh & MATE & CAD & LUCIA & MADE \\
                 & ASR & ASR & ASR & ASR \\
            \midrule
            \textsc{FromReal} & 85.71 & -- & 34.63 & 48.49 \\
            \textsc{Attached} & 30.95 & -- & 29.82 & 17.58 \\
            \textsc{Hollow} & 46.94 & 49.98 & 33.04 & 26.11 \\
            \bottomrule
        \end{tabular}
        }
    \end{subtable}
\end{table}

%% file: discussion.tex
\section{Discussion and Limitations}
\label{sec:discussion}

\myparagraph{Scope of evaluation}
Our study evaluates standard LiDAR-based CP backbones using OPV2V as the primary simulated benchmark and V2X-Real as a complementary real-world benchmark. We additionally collect real-world LiDAR captures of fabricated prototypes. Two extensions are deferred to future work: (i)~advanced CP architectures spanning different modalities, including camera-LiDAR fusion and other multimodal designs, and (ii)~larger-scale real-world experiments with industrially fabricated optimized meshes.

\myparagraph{Mitigation and limitation}
\defensename mitigates trust degradation for benign providers but cannot distinguish benign from malicious uncertainty. A compromised collaborator could self-mask manipulated regions, removing its data from fusion while evading a trust penalty. Such strategic masking or omission is outside our threat model, and distinguishing benign from adversarial inconsistencies remains open.

%% file: conclusion.tex
\section{Conclusion}
\label{sec:conclusion}

We identified an attribution vulnerability in consistency-based collaborative perception defenses: a physical object can induce legitimate cross-view disagreement that the defense misattributes to a benign participant. \attackname realizes this threat without compromising a vehicle or V2X communication, using physically biased shape priors and constrained optimization to suppress victim-view evidence while preserving non-victim observations. Across collaborative perception models and defenses, the resulting disagreement can cause the victim to be downweighted or excluded. We further proposed \defensename, a lightweight self-reflection mechanism that prevents locally uncertain evidence from poisoning collaborator trust.

%% file: appendix.tex
\section{Implementation Details}
\label{app:implementation-details}

We evaluate our attack against SOTA collaborative perception models spanning both late and intermediate fusion paradigms. For late fusion systems, we target PIXOR~\cite{yang2018pixor} and PointPillars~\cite{lang2019pointpillars}. For intermediate fusion architectures, we evaluate Attentive Fusion~\cite{xu2022opv2v} and Where2comm~\cite{hu2022where2comm, xu2022opv2v}, both configured with a PointPillars backbone.

To enable end-to-end optimization against these models, we implement a differentiable LiDAR renderer using M\"oller--Trumbore ray tracing~\cite{moller2005fast} for mesh intersection. We specifically address the non-differentiable components of the target perception models as follows. For PIXOR~\cite{yang2018pixor}, the non-differentiable BEV preprocessor is replaced with a differentiable approximation inspired by SlowLiDAR~\cite{liu2023slowlidar}. For PointPillars~\cite{lang2019pointpillars}, the non-differentiable Voxelization and Scatter operations are substituted with trilinear and $\tanh$ approximations~\cite{cao2021invisible}, and we utilize Gumbel-Softmax~\cite{jang2016categorical} as a differentiable surrogate for its non-differentiable max pooling step. In this way, the entire optimization pipeline is rendered differentiable.

We evaluate four collaborative perception defenses: CAD~\cite{zhang2024data}, MATE~\cite{hallyburton2025security}, LUCIA~\cite{wang2025threat}, and MADE~\cite{zhao2024made}. For CAD, we follow the original implementation and evaluate the CAD-specific \textsc{Hollow} mode described in Section~\ref{subsec:offline-optimization}. For MATE, we follow the original trust-estimation procedure and set the temporal trust period to 10 frames. Unless otherwise stated, we report the 5-track setting as the default MATE result, since the original MATE evaluation uses scenarios with 5--10 vehicles. We further analyze the impact of track count in Section~\ref{subsec:impact-factor}. For LUCIA, we follow the original feature-consistency scoring mechanism and set the compression ratio to 1, which gives LUCIA the finest spatial granularity and avoids weakening it through feature averaging. We study the effect of larger compression ratios in Section~\ref{subsec:impact-factor}. For MADE, we adapt the original method to OPV2V by training the autoencoder using benign CP cases and collecting match-branch calibration scores from benign cases, on both PointPillar AttFusion~\cite{xu2022opv2v} and Where2Comm~\cite{hu2022where2comm}, following its semi-supervised anomaly detection protocol.

For ROBOSAC, we use a three-trial sampling budget, one assumed attacker, a target sampling-success probability of $0.99$, and a box-matching IoU threshold of $0.5$. We force the victim into the first sampled subgroup, while subsequent subgroups follow ROBOSAC's random sampling procedure. Benign and attacked observations use identical sampled subgroups. A pair is successful only when ROBOSAC reaches consensus with a victim-containing subgroup in the benign case and reaches consensus with a victim-excluding subgroup in the attacked case. No-consensus fallback is not counted as attack success. We sweep the box-set consensus threshold from $0.3$ to $0.8$.

For \defensename, we implement the self-reflection step as a lightweight add-on without retraining the base perception or defense models. In our 10-frame evaluation sequences, when the victim observes a local inconsistency with the majority for two consecutive frames, it activates a trust mask on that local inconsistent region from the next frame through the last frame. Downstream defense computation ignores the masked region when computing inconsistency scores, trust updates, or anomaly decisions. For temporal MATE, the mask is applied before the pseudomeasurement update in each masked frame.

For the scenario-sensitivity analysis in Figure~\ref{fig:selected-case-delta-comparison}, we additionally form two 50-scenario critical subsets from the 430-scenario benchmark. The ASR-oriented subset favors cases with larger victim--non-victim pose separation, suitable non-victim view angles, stronger occlusion, and more CAVs, which make cross-view inconsistency easier to attribute to the victim. The AP-oriented subset favors cases where the victim has a larger unique contribution to collaborative perception, so downweighting or excluding it causes greater utility loss. These subsets operationalize the scenario assessment in Section~\ref{subsec:online-deployment}. All main tables still report the full benchmark.

For the deployment-robustness experiment in Table~\ref{tab:deployment-robustness}, each frame applies a yaw offset around the vertical axis sampled uniformly from $\ang{-10}$ to $\ang{10}$ to the adversarial object and its ground-truth box. We reuse the OPV2V scenarios, meshes, and evaluation settings of the precisely aligned experiment, and evaluate MATE, CAD, LUCIA, and MADE without \defensename.

Optimization runs for 300 epochs using Adam with a learning rate of 0.01. We set the vulnerable view cone $\Theta_v$ to $\ang{5}$ behind the object and sample non-vulnerable views $\Theta_n$ with at least $\ang{20}$ angular separation from $\Theta_v$. The loss weights are set to $\alpha=1$, $\beta=1$, and $\lambda=0.001$ for Laplacian regularization. The \textsc{Hollow} mode applies per-vertex $\tanh$ projection to bound deformation (0.1~m), while \textsc{Attached} constrains maximum translation and overall mesh size to preserve stealthiness. The global translation is constrained to a range within 0.3~m of the rear bumper. The size of the initial \textsc{Hollow} mesh is similar to a normal car, measuring 4.4~m $\times$ 1.6~m $\times$ 1.6~m. LiDAR simulation in the synthetic data generation matches OPV2V specifications (64-channel, 70~m range). The feature number of PointPillars is set to 128 to allow for larger and more effective perturbations.

\section{Optimized Adversarial Objects}
\label{app:optimized-meshes}

\begin{figure}[t]
    \centering
    \begin{subfigure}[t]{0.32\columnwidth}
        \centering
        \includegraphics[width=\columnwidth]{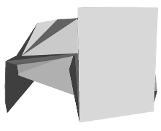} \\ \vspace{2pt}
        \includegraphics[width=\columnwidth]{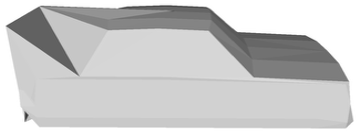}
        \caption{\small \textsc{FromReal}}
        \label{fig:adv-real}
    \end{subfigure}
    \hfill
    \begin{subfigure}[t]{0.32\columnwidth}
        \centering
        \includegraphics[width=\columnwidth]{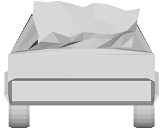} \\ \vspace{2pt}
        \includegraphics[width=\columnwidth]{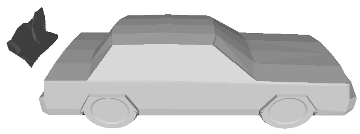}
        \caption{\small \textsc{Attached}}
        \label{fig:adv-attached}
    \end{subfigure}
    \hfill
    \begin{subfigure}[t]{0.32\columnwidth}
        \centering
        \includegraphics[width=\columnwidth]{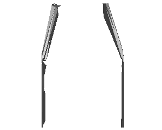} \\ \vspace{2pt}
        \includegraphics[width=\columnwidth]{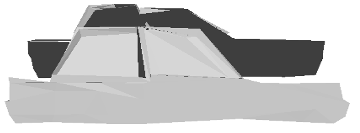}
        \caption{\small \textsc{Hollow}}
        \label{fig:adv-hollow}
    \end{subfigure}

    \caption{Representative adversarial objects optimized on PointPillars late fusion. The top and bottom rows show the back and oblique top-side views, respectively.}
    \label{fig:adversarial-mesh-comparison}
\end{figure}

The optimized meshes retain the physical bias of their initial shape priors. \textsc{FromReal} remains vehicle-shaped, \textsc{Attached} confines the perturbation to a compact rear-mounted object, and \textsc{Hollow} preserves its open-channel structure (Figure~\ref{fig:adversarial-mesh-comparison}).

